\documentclass[11pt]{article}
\usepackage[preprint]{acl}

\usepackage{fontspec}
\newfontfamily\banglafont{NotoSansBengali-Regular.ttf}[
  Path=fonts/,
  BoldFont=NotoSansBengali-Bold.ttf,
  Script=Bengali
]
\newcommand{\bn}[1]{{\banglafont #1}}
\usepackage{tikz}
\usetikzlibrary{positioning, arrows.meta, calc, shapes.geometric}
\usepackage{pgfplots}
\pgfplotsset{compat=1.18}
\usepgfplotslibrary{fillbetween}
\usepackage{booktabs}
\usepackage{array}   
\usepackage{multirow}
\usepackage{xcolor}
\usepackage{microtype}
\usepackage{graphicx}
\usepackage{amsmath}

\usepackage{newtxmath}
\usepackage{xspace}
\usepackage{pifont}
\usepackage{enumitem}
\usepackage{algorithm}
\usepackage{algpseudocode}
\definecolor{v3red}{RGB}{0,0,0}
\newcommand{\new}[1]{#1}
\newcommand{\sys}{BanglaBox\xspace}
\newcommand{\cmark}{\ding{51}}
\newcommand{\xmark}{\ding{55}}

\makeatletter
\def\thebibliography#1{\vskip\parskip%
\vskip\baselineskip%
\def\baselinestretch{1}%
\ifx\@currsize\normalsize\@normalsize\else\@currsize\fi%
\vskip-\parskip%
\vskip-\baselineskip%
\section*{References\@mkboth
 {References}{References}}\list
 {}{\setlength{\labelwidth}{0pt}\setlength{\leftmargin}{\parindent}
 \setlength{\itemindent}{-\parindent}%
 \setlength{\itemsep}{0pt}\setlength{\parsep}{0pt}\setlength{\topsep}{2pt}}
 \def\newblock{\hskip .11em plus .33em minus -.07em}
 \sloppy\clubpenalty4000\widowpenalty4000
 \sfcode`\.=1000\relax}

\makeatother

\title{\sys: A Phonetically-Balanced Corpus and Data-Efficient Foundation-Model Adaptation for Bangla Text-to-Speech with Zero-Shot Voice Cloning}

\author{
  Emtiaz Uddin Ahmed \\
  \texttt{emtiaz4546@gmail.com}
  \And
  Araf Mahmud
  \And
  Sajib Hossain
  \AND
  Tarikul Islam Tamiti
  \And
  Sajid Fardin Dipto
  \And
  Anomadarshi Barua
}

\hypersetup{
  pdftitle={BanglaBox: A Phonetically-Balanced Corpus and Data-Efficient Foundation-Model Adaptation for Bangla Text-to-Speech with Zero-Shot Voice Cloning},
  pdfauthor={Emtiaz Uddin Ahmed; Araf Mahmud; Sajib Hossain; Tarikul Islam Tamiti; Sajid Fardin Dipto; Anomadarshi Barua},
  pdfsubject={Bangla text-to-speech, corpus construction, and zero-shot voice cloning},
  pdfkeywords={Bangla, Bengali, text-to-speech, TTS, voice cloning, speech corpus}
}

\AtBeginDocument{%
  \setlength{\abovecaptionskip}{10pt}%
  \setlength{\belowcaptionskip}{0pt}%
  \setlength{\textfloatsep}{10pt plus 2pt minus 5pt}%
  \setlength{\floatsep}{6pt plus 1pt minus 1pt}%
  \setlength{\intextsep}{0pt plus 1pt minus 1pt}%
  \setlength{\dbltextfloatsep}{14pt plus 2pt minus 2pt}%
  \setlength{\dblfloatsep}{0pt plus 1pt minus 1pt}%
  \setlength{\abovedisplayskip}{4pt plus 2pt minus 2pt}%
  \setlength{\belowdisplayskip}{4pt plus 2pt minus 2pt}%
  \setlength{\abovedisplayshortskip}{2pt}%
  \setlength{\belowdisplayshortskip}{2pt}%
}

\makeatletter
\renewcommand\section{\@startsection{section}{1}{\z@}{-2.0ex}{1.5ex}{\large\bfseries\raggedright}}
\renewcommand\subsection{\@startsection{subsection}{2}{\z@}{-1.8ex}{0.8ex}{\normalsize\bfseries\raggedright}}
\makeatother

\begin{document}
\maketitle

\begin{abstract}
\footnote{\new{All artifacts are public and unconditional, not contingent on acceptance: corpus (CC BY 4.0), model weights, LoRA adapter, tokenizer extension, text-normalisation scripts, training and inference code (MIT), and the complete evaluation materials including every per-rating record. \url{https://huggingface.co/Banglabox}}}

We present a recipe for adapting English-pretrained autoregressive TTS foundation models to underrepresented languages, demonstrated on Bangladeshi Bangla.
Existing Bangla TTS corpora are small and single-speaker, and to our knowledge no open zero-shot voice-cloning system is available for the Bangladeshi register.
We contribute a phonetically- and gender-balanced two-tier Bangladeshi Bangla corpus balanced via a tiered Jensen--Shannon divergence objective over conjunct clusters (\textit{juktakkhor}), together with 3 fine-tuning changes---a \emph{merge-consistent tokenizer extension}, Bangla text normalization, and a prompt-masked dual-loss objective---that preserve zero-shot cloning across the language switch.
Our \emph{BanglaEval} protocol applies Wilcoxon signed-rank tests with Bonferroni correction over native-speaker ratings.
\sys{} attains near-natural Naturalness, outperforms commercial and open-source baselines on Naturalness, Speaker Similarity, and Clarity, and reaches speaker similarity comparable to prior few-shot results while using  $\sim7\times$ less Bangla fine-tuning audio.
 \new{We further validate the recipe beyond our own test split: on a seven-category stress set of difficult real-world text, on the public BnTTS evaluation benchmarks, and on naturally occurring Bangla that no language model wrote. All artifacts --- corpus, weights, code and the complete evaluation materials --- are released publicly and unconditionally.}
\end{abstract}

\section{Introduction}
\label{sec:intro}

\begin{figure*}[!t]
\centering
\resizebox{0.86\textwidth}{!}{%
\begin{tikzpicture}[
    every node/.style={font=\scriptsize, align=center},
    src/.style       ={draw, rounded corners=4pt, fill=green!12,  draw=green!55!black, line width=0.6pt, text width=3.2cm, minimum height=1.0cm, inner sep=4pt, font=\small},
    proc/.style      ={draw, rounded corners=4pt, fill=blue!8,    draw=blue!55,        line width=0.6pt, text width=2.4cm, minimum height=1.0cm, inner sep=4pt},
    llmbox/.style    ={draw, rounded corners=4pt, fill=purple!8,  draw=purple!55,      line width=0.6pt, text width=2.6cm, minimum height=1.0cm, inner sep=4pt},
    gate/.style      ={draw, fill=orange!12, draw=orange!70,      line width=0.6pt, text width=3.0cm, minimum height=1.0cm, inner sep=4pt},
    review/.style    ={draw, ellipse, fill=red!10,  draw=red!55,  line width=0.6pt, text width=2.0cm, inner sep=4pt, minimum height=1.0cm},
    decide/.style    ={draw, diamond, aspect=2, fill=yellow!15, draw=olive,  line width=0.6pt, inner sep=2pt, minimum height=0.9cm},
    outbox/.style    ={draw, rounded corners=4pt, fill=teal!12, draw=teal!60,  line width=0.7pt, text width=2.2cm, minimum height=1.0cm, inner sep=4pt},
    arr/.style       ={->, >=stealth, line width=0.7pt, color=black!75},
    sidearr/.style   ={->, >=stealth, line width=0.7pt, color=black!50, dashed},
]

\node[src]    (prompt)  at (-0.4, 0) {Coverage-feedback prompts \\ \textit{(juktakkhor,} \\ \textit{domain, gender)}};
\node[llmbox] (llm)     at (3.8,  0) {LLM script gen \\ {\small GPT-4o-mini ($\sim$90\%)} \\ {\small GPT-4o ($\sim$7\%)} \\ {\small Claude Opus ($\sim$3\%)}};
\node[gate]   (gates)   at (7.8,  0) {9 automated quality gates \\ \textit{length, TTR, dedup, lang-ID, perplexity, phonetic novelty, script, semiotic, gap-unit}};
\node[review] (humrev)  at (11.6, 0) {Two-pass human review \\ \textit{linguist + ethics reviewer}};
\node[outbox] (tierA)   at (15.0, 0) {\textbf{Tier A scripts} \\ \textit{studio-ready}};

\draw[arr] (prompt) -- (llm);
\draw[arr] (llm)    -- (gates);
\draw[arr] (gates)  -- (humrev);
\draw[arr] (humrev) -- (tierA);

\draw[sidearr] (gates.south)  -- ++(0,-0.35) -| (llm.south);
\draw[sidearr] (humrev.south) -- ++(0,-0.70) -| (llm.south east);
\node[font=\tiny, color=black!55, above=1pt] at (5.8, -0.85) {reject};

\node[proc] (studio) at (15.0, -2.4) {Dhaka studio recording \\ \textit{48\,kHz/24-bit, SNR $\geq$ 44\,dB}};
\draw[arr] (tierA) -- (studio);

\node[src]    (hfsrc)   at (-0.4, -2.4) {Public HF Bangla TTS datasets \\ \textit{(39 speakers, 247\,h init.)}};
\node[gate]   (filter)  at (3.8,  -2.4) {Speaker filter \\ \textit{SNR $\geq$ 35\,dB + ablation-driven curation}};
\node[decide] (drop)    at (7.8,  -2.4) {Keep?};
\node[outbox] (tierB)   at (11.6, -2.4) {\textbf{Tier B audio} \\ \textit{23 speakers (16 dropped)}};

\draw[arr] (hfsrc)  -- (filter);
\draw[arr] (filter) -- (drop);
\draw[arr] (drop)   -- node[above, font=\tiny]{yes} (tierB);
\draw[sidearr] (drop.south) -- node[right, font=\tiny]{no} ++(0, -0.5);

\node[outbox, text width=11.6cm, minimum height=1.3cm, font=\small] (corpus) at (10.4, -4.2)
    {\textbf{BanglaBox-520 corpus} \quad 520.3\,h \quad 36 speakers \quad 7 domains \\[2pt]
     juktakkhor-aware tiered JSD balance \quad speaker-disjoint test split};

\draw[arr] (tierB.south) -- (tierB.south |- corpus.north);

\draw[arr] (studio.south) -- (studio.south |- corpus.north);

\node[draw, dashed, rounded corners=2pt, fill=gray!8, draw=gray!50, inner sep=6pt, font=\scriptsize, text width=5.8cm, align=center]
    (eq) at (0.95, -4.2)
    {Tiered JSD balance: \\[2pt]
    \resizebox{\linewidth}{!}{$\min \sum_{i=1}^{4} w_i \, \mathrm{JSD}(p_i^{\text{corpus}} \,\|\, p_i^{\text{target}})$} \\[2pt]
    over phone / diphone / triphone / juktakkhor};

\draw[sidearr] (eq.east) -- (corpus.west);

\end{tikzpicture}%
}
\caption{Two-tier corpus construction (\S\ref{sec:dataset}). Tier A passes 9 quality gates (thresholds: Table~\ref{tab:gates}, App.~\ref{app:gates}) and two-pass human review before studio recording; Tier B is filtered by signal-to-noise ratio (SNR) and an ablation check (16/39 speakers dropped). Tiers merge under the tiered Jensen--Shannon divergence (JSD) objective (Eq.~\ref{eq:jsd}).}
\label{fig:dataset_pipeline}
\end{figure*}
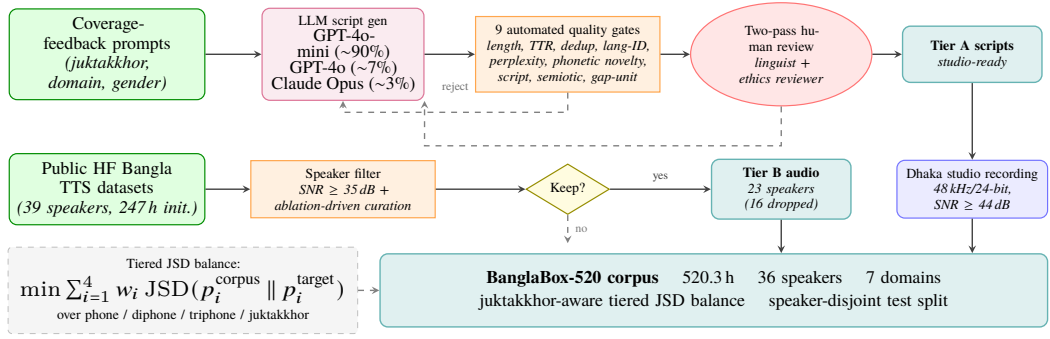

Bangla (Bengali), spoken by over 230 million people \cite{ethnologue2024}, remains severely underrepresented in neural speech synthesis.
Existing systems share one limitation: HMM \cite{mukherjee2012hmm}, LSTM-RNN \cite{gutkin2016}, VITS \cite{vits_bangla2025} and DNN-based \cite{sust_dnn_tts} approaches are all bound to small \emph{single-speaker} corpora, precluding voice cloning, cross-speaker generalization and multi-domain coverage.
The SUST TTS Corpus \cite{sust_tts_corpus}, the largest prior balanced Bangla resource, is single-speaker, and prior balance design relies on phone/diphone coverage \cite{sust_diphone,sust_duration,sust_phoneme} without modelling the conjunct-cluster (\textit{juktakkhor}) inventory that drives Bangla naturalness.
BnTTS \cite{bntts2025}, the recent state of the art, fine-tunes an XTTS \cite{casanova2024xtts} backbone on 3.85k\,h of Bangla speech. However, its weights are not public, its evaluation lacks significance testing, and its data requirement is prohibitive for low-resource follow-up. Zero-shot TTS in codec-language \cite{wang2023valle,kharitonov2023speak}, latent-diffusion \cite{shen2024naturalspeech2,li2023styletts2}, and flow-matching \cite{le2023voicebox} families achieves strong English results but is unevaluated on Bangla. Moreover, the open Indic-language model IndicF5 \cite{indicf5} reaches only SECS score of $0.37$ and fine-tuned YourTTS \cite{casanova2022yourtts} reaches $0.44$, both far below human.

Pretrained TTS foundation models offer a path forward.
Chatterbox TTS \cite{chatterboxtts2025} is an MIT-licensed Llama-style autoregressive transformer with built-in zero-shot cloning from a reference clip.\footnote{\small Chatterbox has not undergone formal peer review; we cite the released model and its Hugging Face model card.}
Adapting it to Bangla means handling juktakkhor, context-dependent vowel diacritics and honorific abbreviations absent from standard toolchains \cite{bnnumerizer,bnunicodenormalizer,ghosh2024textnorm}, while preserving cloning across the language switch.
To our knowledge, no prior work adapts Chatterbox to Bangla, nor formalizes merge-consistent vocabulary extension when importing new scripts into autoregressive codec language models. Commercial services -- Google, Azure, Narakeet~\cite{narakeet} -- provide Bangla synthesis. Therefore, we include them as baselines.

We address these gaps with \sys, a documented fine-tuning recipe for Bangladeshi Bangla (Dhaka register).
\sys{} leads all 5 baselines on every perceptual metric ($p{<}0.01$, Wilcoxon with Bonferroni correction).
It beats Google by $+6.8\%$ on Naturalness and IndicF5 by $+17.2\%$, and nearly doubles IndicF5's cloning fidelity ($0.627$ vs.\ $0.37$ SECS; commercial systems expose no speaker adaptation). Per-system margins are in Table~\ref{tab:main_results}.
\sys{} reaches speaker similarity comparable to prior few-shot Bangla results (indicative; encoders and model families differ, \S\ref{sec:experiments}), trained on $520.3$\,h of Bangla audio against the $3.85$k\,h of prior work, and sits within 0.20 MOS of the $4.76$ human ceiling. In summary, our contributions are:

\textbf{(i) Corpus.} A $520.3$\,h, $36$-speaker, $7$-domain Bangladeshi Bangla corpus, phonetically- and gender-balanced via a novel \emph{juktakkhor-aware tiered Jensen--Shannon divergence (JSD)} objective (\S\ref{sec:dataset}). \textbf{(ii) Three-part adaptation.} A merge-consistent tokenizer extension, Bangla text normalization, and a prompt-masked dual-loss objective \cite{wang2023valle}, each independently ablated (\S\ref{sec:method}, \S\ref{sec:experiments}). \textbf{(iii) BanglaEval.} A reproducible protocol with $24$ native raters and Wilcoxon signed-rank testing under Bonferroni correction --- stronger statistical grounding than any prior Bangla TTS evaluation --- plus a chunked long-form inference pipeline. \textbf{(iv) System.} The first Chatterbox-based zero-shot Bangla voice-cloning system; misuse risks are addressed in the Ethics Statement.
\new{\textbf{(v) Validation.} A validation suite beyond the internal split --- a seven-category stress test, the public BnTTS benchmarks, a natural-text register control, a same-corpus XTTS-v2 subjective comparison and a three-seed matched-hours re-run --- with a full public artifact release.}

\section{Dataset construction}
\label{sec:dataset}

\subsection{Two-tier corpus design}

We construct a 520.3-hour Bangladeshi Bangla TTS corpus from 36 speakers in 2 tiers (Fig.~\ref{fig:dataset_pipeline}).
\textbf{Tier A} (367.3\,h, 13 speakers, $\sim$28.3\,h/speaker) is studio-recorded in Dhaka via a coverage-feedback loop: LLM-generated scripts (GPT-4o-mini ${\sim}90\%$, GPT-4o ${\sim}7\%$ for rare-conjunct gap-targeting, Claude Opus ${\sim}3\%$ for dialogue) pass 9 automated quality gates (thresholds in Table~\ref{tab:gates}, App.~\ref{app:gates}) and a two-pass human review before professional recording at $\ge$44\,dB SNR (48\,kHz archive, 24\,kHz working copy).
\textbf{Tier B} (153.0\,h, 23 speakers, $\sim$6.7\,h/speaker) is curated from 7 public Bangla speech datasets (6 on Hugging Face, one from the Bengali.AI portal; licenses and per-dataset retention in Table~\ref{tab:tierb_sources}, App.~\ref{app:tierb}), retaining only speakers that pass an SNR filter and an ablation-driven check on downstream quality; 16 of 39 initial speakers (94\,h of 247\,h) were dropped.
Both tiers closely match a 7-domain distribution and target gender balance.

\begin{table}[t]
\centering
\small
\begin{tabular}{lrrr}
\toprule
\textbf{Partition} & \textbf{Spk.} & \textbf{Hours} & \textbf{Utts.} \\
\midrule
\multicolumn{4}{l}{\emph{By tier}} \\
Tier A (studio)   & 13 & 367.3 & $\sim$220{,}000 \\
Tier B (curated)  & 23 & 153.0 & $\sim$92{,}000  \\
\midrule
\multicolumn{4}{l}{\emph{By split (speaker-disjoint test)}} \\
Train (11A + 20B) & 31 & $\sim$422 & $\sim$253{,}000 \\
Dev (within-spk)  & 31 & $\sim$22  & $\sim$13{,}000  \\
Test (2A + 3B)    &  5 & $\sim$76  & $\sim$46{,}000  \\
\midrule
\textbf{Total} & \textbf{36} & \textbf{520.3} & \textbf{$\sim$312{,}000} \\
\bottomrule
\end{tabular}
\caption{Corpus statistics by tier and split ($\bar{d}{=}6.0$\,s mean utterance duration).}
\label{tab:dataset_stats}
\end{table}

\begin{table}[t]
\centering
\footnotesize
\setlength{\tabcolsep}{2.5pt}
\resizebox{\columnwidth}{!}{%
\begin{tabular}{@{}llrrccc@{}}
\toprule
\textbf{Dataset} & \textbf{Yr} & \textbf{Hrs} & \textbf{Spk} & \textbf{MS} & \textbf{Cl} & \textbf{Dom.} \\
\midrule
SUBESCO          & 2021 &   7.7 &  20 & \cmark & \xmark & Emot. \\
SHRUTI           & 2011 &    22 &  34 & \cmark & \xmark & Read \\
SUST TTS         & 2021 &    30 &   1 & \xmark & \xmark & Read \\
CommonVoice (v9) & 2022 &    56 &  -- & \cmark & \xmark & Crowd \\
\midrule
\multicolumn{7}{@{}l}{\new{\emph{Widely used non-English TTS corpora}}}\\
\new{LJSpeech}   & \new{2017} & \new{24}  & \new{1}    & \new{\xmark} & \new{\xmark} & \new{Audiobook} \\
\new{VCTK}       & \new{2019} & \new{44}  & \new{110}  & \new{\cmark} & \new{\xmark} & \new{Read} \\
\new{LibriTTS}   & \new{2019} & \new{585} & \new{2456} & \new{\cmark} & \new{\xmark} & \new{Audiobook} \\
\new{AISHELL-3}  & \new{2020} & \new{85}  & \new{218}  & \new{\cmark} & \new{\xmark} & \new{Read} \\
\new{CSS10}      & \new{2019} & \new{140} & \new{10}   & \new{\xmark} & \new{\xmark} & \new{Audiobook} \\
\midrule
\textbf{Ours (\sys)} & \textbf{2025} & \textbf{520.3} & \textbf{36} & \cmark & \cmark & \textbf{7 dom.} \\
\bottomrule
\end{tabular}%
}
\caption{Comparison with existing Bangla speech datasets (MS = multi-speaker; Cl = cloning). \sys{} is $\sim 9\times$ Common Voice Bengali v9 and $\sim 17\times$ the SUST TTS Corpus. \new{Widely used non-English corpora are included for scale; none applies phonetic balancing, whereas \sys{} balances phone, diphone, triphone and juktakkhor distributions.}}
\label{tab:dataset_compare}
\vspace{-0.3em}    
\end{table}

\begin{figure}[t]
\noindent\includegraphics[width=\columnwidth]{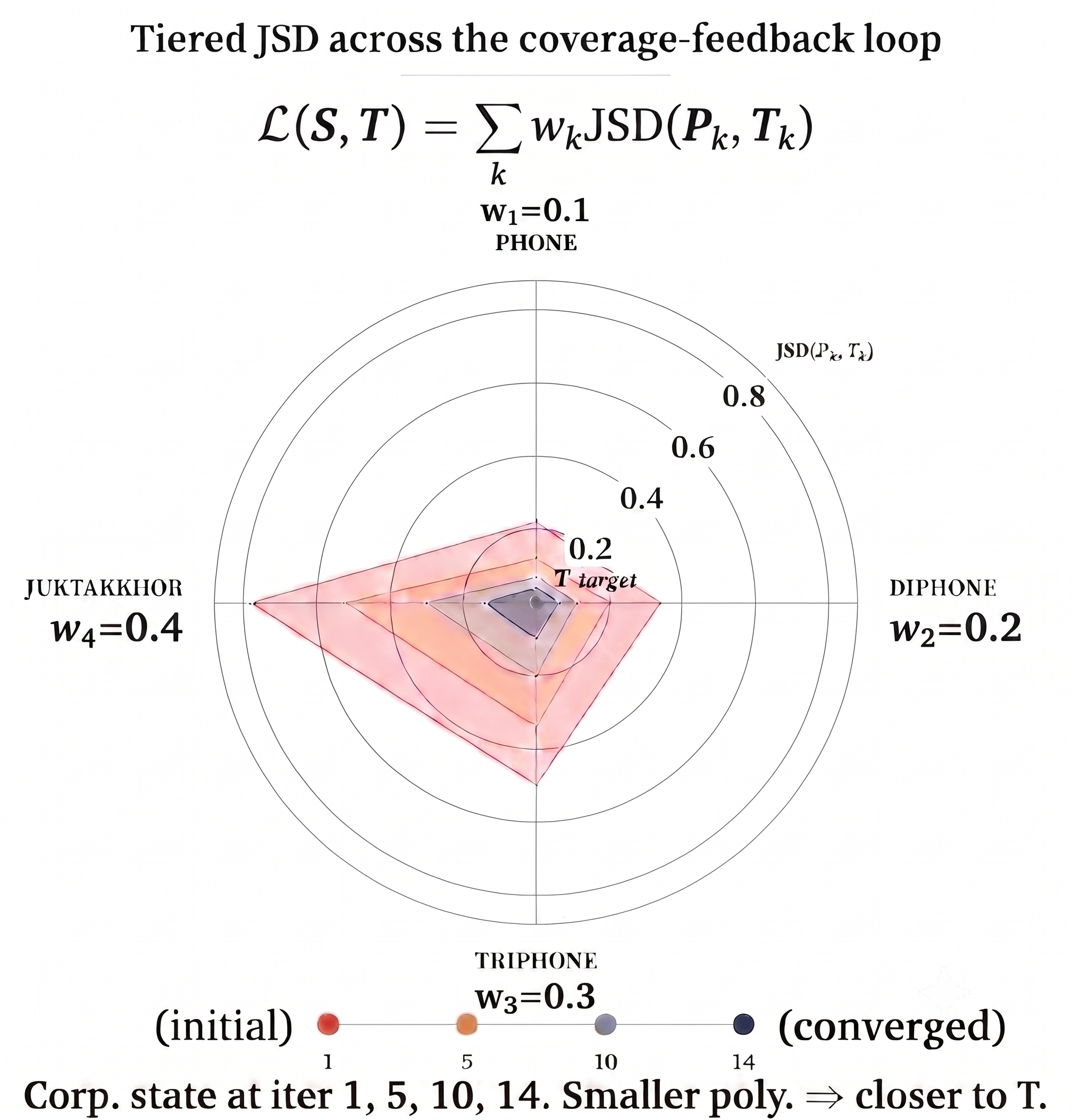}
\caption{Tiered JSD across coverage-feedback iterations 1, 5, 10, 14; smaller polygon $\Rightarrow$ closer to target~$T$.}
\label{fig:jsd_loop}
\vspace{-0.10em}   
\end{figure}

\subsection{Tiered JSD balance objective}

Tier A script selection minimizes a \emph{tiered Jensen--Shannon divergence} loss over phone, diphone, triphone, and juktakkhor histograms:

\begin{equation}
\small
\mathcal{L}(S,T) = \!\! \sum_{k\in\{p,d,t,j\}} \!\! w_k \cdot \text{JSD}(P_k, T_k),
\label{eq:jsd}
\end{equation}

where $P_k$ is the unit histogram of the selected script set $S$ at tier $k$, over Bangla's phone, diphone, triphone, and conjunct (\textit{juktakkhor}) inventories \cite{sust_tts_corpus,sust_diphone,sust_duration,sust_phoneme}. The $\sim$300-form juktakkhor inventory is extended via the coverage-feedback loop.
The weights $(w_p,w_d,w_t,w_j){=}(0.1,0.2,0.3,0.4)$ are not fitted. They follow the simplest increasing scheme summing to one ($w_k{=}k/10$) after ranking the tiers from easiest to hardest to cover with natural text --- phones are few and frequent, diphones common in ordinary text, triphones need targeted text, and juktakkhor are rarest and not reducible to letter sequences (\bn{ক্ষ}, \bn{জ্ঞ} are pronounced differently from their components) --- a ranking mirrored in each tier's convergence order in Fig.~\ref{fig:jsd_loop} (weight-sensitivity analysis: App.~\ref{app:weight_sens}).
The target $T_k$ for tier $k$ is a Jelinek--Mercer-style linear interpolation between the natural unit frequencies of the candidate pool and a uniform distribution over the full tier-$k$ inventory, $T_k = \lambda\, P^{\text{nat}}_k + (1-\lambda)\, U_k$ with $\lambda{=}0.7$: the natural part keeps the selected text realistic, while the uniform part keeps a non-zero target on every unit --- including the rarest conjuncts --- so the objective can never be minimized while ignoring rare units.
Selection is additionally informed by Bangla lexicon and loanword literature \cite{lexicon_bangla,loanwords_bangla,collab_lexicon_bangla,news_corpus_bangla,semantic_lexicon_bangla}.
Algorithm~\ref{alg:jsd} (App.~\ref{app:algo}) summarizes the full selection loop --- candidate generation, gating, review, and greedy JSD-driven admission.
Tier A achieves $\ge$99\% phone, $\ge$93\% diphone, $\ge$80\% phonotactically-valid triphone, and $\ge$85\% juktakkhor coverage (top-150 juktakkhor: $\ge$97\%); these figures are \emph{achieved} outcomes of the loop, not preset stopping thresholds. Fig.~\ref{fig:jsd_loop} visualizes the per-tier JSD shrinkage across the 14-iteration coverage-feedback loop, with phone and diphone tiers converging fastest and the juktakkhor tier converging last, reflecting its inherent sparsity in Bangla orthography and phonotactics.

\begin{figure*}[!t]
\centering
\definecolor{clrA}{RGB}{83,74,183}
\definecolor{clrB}{RGB}{29,158,117}
\definecolor{clrC}{RGB}{216,90,48}
\definecolor{clrD}{RGB}{212,83,126}
\definecolor{clrE}{RGB}{55,138,221}

\begin{minipage}[t]{0.32\textwidth}
\centering
\begin{tikzpicture}
\begin{axis}[
    ybar,
    width=\textwidth,
    height=4.5cm,
    ylabel={\footnotesize Utterances},
    ylabel style={font=\footnotesize},
    xlabel={\footnotesize Utterance duration (s)},
    xlabel style={font=\footnotesize},
    ytick={0,30000,60000,90000,120000},
    yticklabels={0,30k,60k,90k,120k},
    yticklabel style={font=\scriptsize},
    ymin=0, ymax=135000,
    symbolic x coords={2-4s,4-6s,6-8s,8-10s,10s+},
    xtick=data,
    xticklabels={2--4,4--6,6--8,8--10,10+},
    xticklabel style={font=\footnotesize},
    bar width=9pt,
    enlarge x limits=0.18,
    ymajorgrids=true,
    grid style={gray!12},
    nodes near coords,
    nodes near coords style={font=\scriptsize, color=black},
    point meta=explicit symbolic,
]
\addplot[draw=none, fill=clrA, fill opacity=0.3] coordinates {
    (2-4s,107052) [107k]
    (4-6s,114339) []
    (6-8s,13009)  [13k]
    (8-10s,3179)  [3k]
    (10s+,3814)   [4k]
};
\addplot[draw=none, fill=clrB, fill opacity=0.9] coordinates {
    (4-6s,114339) [114k]
};
\end{axis}
\end{tikzpicture}
\par\vspace{-7pt}
{\footnotesize (a) Duration distribution}
\end{minipage}%
\hfill
\begin{minipage}[t]{0.32\textwidth}
\centering
\begin{tikzpicture}
\begin{axis}[
    ybar,
    width=\textwidth,
    height=4.5cm,
    ylabel={\footnotesize Utterances},
    ylabel style={font=\footnotesize},
    ytick={0,25000,50000,75000,100000},
    yticklabels={0,25k,50k,75k,100k},
    yticklabel style={font=\scriptsize},
    ymin=0, ymax=120000,
    symbolic x coords={1-4,5-6,7-8,9-10,11-14,15+},
    xtick=data,
    xticklabel style={font=\scriptsize},
    xlabel={\footnotesize Words per utterance},
    xlabel style={font=\footnotesize},
    bar width=8pt,
    enlarge x limits=0.12,
    ymajorgrids=true,
    grid style={gray!12},
    nodes near coords,
    nodes near coords style={font=\scriptsize, color=black},
    point meta=explicit symbolic,
]
\addplot[draw=none, fill=clrC, fill opacity=0.3] coordinates {
    (1-4,190)     [0.2k]
    (5-6,82894)   [83k]
    (7-8,103713)  []
    (9-10,35575)  [36k]
    (11-14,15063) [15k]
    (15+,3739)    [4k]
};
\addplot[draw=none, fill=clrC, fill opacity=0.9] coordinates {
    (7-8,103713) [104k]
};
\end{axis}
\end{tikzpicture}
\par\vspace{-7pt}
{\footnotesize (b) Word count distribution}
\end{minipage}%
\hfill
\begin{minipage}[t]{0.32\textwidth}
\centering
\begin{tikzpicture}
\begin{axis}[
    width=\textwidth,
    height=4.5cm,
    ylabel={\footnotesize Vocabulary},
    ylabel style={font=\footnotesize},
    xlabel={\footnotesize Utterances ($\times$10\textsuperscript{3})},
    xlabel style={font=\footnotesize},
    xticklabel style={font=\footnotesize},
    yticklabel style={font=\scriptsize},
    xmin=0, xmax=260,
    ymin=0, ymax=80,
    xtick={0,50,100,150,200,250},
    ytick={0,20,40,60,80},
    yticklabels={0,20k,40k,60k,80k},
    grid=major,
    grid style={gray!12},
]
\addplot[name path=vocab, very thick, color=clrD, mark=o, mark size=2pt, mark options={fill=clrD}] coordinates {
    (10,11.596) (50,30.392) (100,44.275) (150,56.010) (200,64.487) (241,68.653)
};
\addplot[name path=zero, draw=none] coordinates {(10,0) (241,0)};
\addplot[clrD, fill opacity=0.12] fill between[of=vocab and zero];
\node[font=\scriptsize, clrD, anchor=south, fill=white, inner sep=1pt] at (axis cs:220,69) {\textbf{68.7k}};
\end{axis}
\end{tikzpicture}
\par\vspace{-7pt}
{\footnotesize (c) Vocabulary growth}
\end{minipage}

\vspace{3pt}

\begin{tikzpicture}
\node[fill=clrA!10, draw=clrA!30, rounded corners=4pt, inner sep=4pt, font=\small] at (0,0) {\textcolor{clrA}{\textbf{$\sim$312k}} utterances};
\node[fill=clrB!10, draw=clrB!30, rounded corners=4pt, inner sep=4pt, font=\small] at (3.5,0) {\textcolor{clrB}{\textbf{520.3}} hours};
\node[fill=clrC!10, draw=clrC!30, rounded corners=4pt, inner sep=4pt, font=\small] at (6.5,0) {\textcolor{clrC}{\textbf{36}} speakers};
\node[fill=clrD!10, draw=clrD!30, rounded corners=4pt, inner sep=4pt, font=\small] at (9.5,0) {\textcolor{clrD}{\textbf{7}} domains};
\node[fill=clrE!10, draw=clrE!30, rounded corners=4pt, inner sep=4pt, font=\small] at (12.5,0) {\textcolor{clrE}{\textbf{220}} chars};
\end{tikzpicture}
\vspace{-3pt}

\caption{Corpus analysis. \textbf{(a)}~Utterance durations peak at 4--6\,s (green). \textbf{(b)}~Word counts peak at 7--8 words (orange). \textbf{(c)}~Vocabulary follows sub-linear Heaps' law growth. Histograms in (a) and (b) cover the $\sim$241\,k utterances of duration $\geq$2\,s; bottom pills summarize statistics for the full 520.3\,h / $\sim$312\,k-utterance corpus.}
\label{fig:corpus_analysis}
\vspace{-3pt}
\end{figure*}
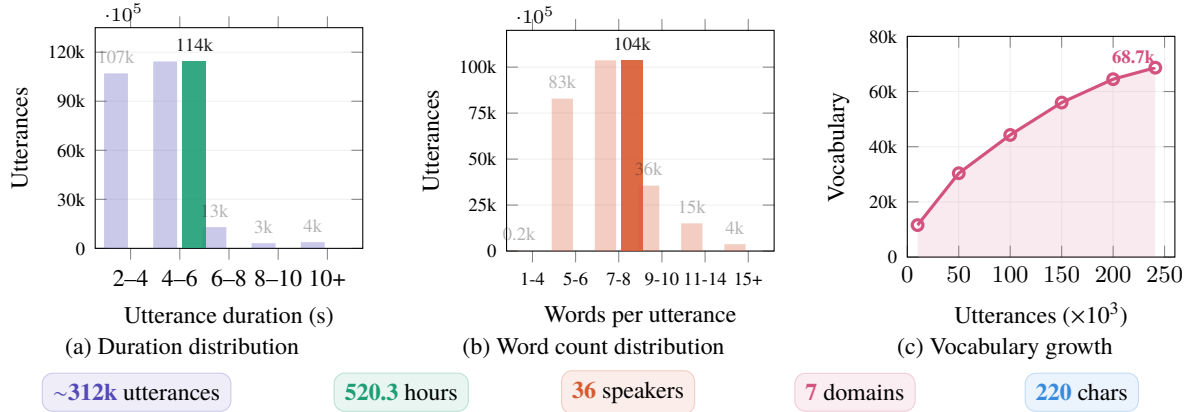

\subsection{Statistics and splits}

Table~\ref{tab:dataset_stats} reports $\sim$312k utterances at $\sim$6.0\,s mean across 7 domains (news, customer care, teaching, healthcare, e-commerce, finance, IT) and a speaker-disjoint train/dev/test split with 5 held-out speakers ($\sim$76\,h) for zero-shot evaluation. Fig.~\ref{fig:corpus_analysis} shows the utterance-duration/word-count/vocabulary distributions.
Table~\ref{tab:dataset_compare} compares \sys against public Bangla speech datasets. Table~\ref{tab:domain_dist} reports Tier A's domain distribution.

\subsection{Tier A quality assurance}

Each studio session is acoustically verified ($\geq$44\,dB SNR, clipping check, breath/lip-smack detection) before alignment; failed segments are re-recorded in the same session. A Bangla-speaking linguist audits a $5\%$ random sample/speaker for pronunciation correctness and prosodic naturalness, with $\geq$99\% segment-level acceptance. The coverage-feedback loop ran for 14 weekly iterations, lifting juktakkhor coverage from $\sim$70\% (week~1) to the $\geq$85\% reported above; measured per-gate rejection counts and rates from the instrumented generation run are reported in Table~\ref{tab:gate_rates}, with each gate's threshold and the failure it blocks in Table~\ref{tab:gates} (App.~\ref{app:gates}).
\section{Methodology}
\label{sec:method}

\begin{figure*}[!t]
\centering
\resizebox{0.86\textwidth}{!}{%
\begin{tikzpicture}[
    every node/.style ={font=\footnotesize, align=center},
    box/.style        ={draw, rounded corners=4pt, align=center, font=\footnotesize, line width=0.6pt, inner sep=5pt, text width=3.5cm, minimum height=1.6cm},
    frozen/.style     ={box, fill=blue!10,   draw=blue!55,   line width=0.7pt},
    s1/.style         ={box, fill=yellow!30, draw=orange!85, line width=0.9pt},
    s2/.style         ={box, fill=orange!25, draw=orange!85, line width=0.9pt},
    s3/.style         ={box, fill=red!18,    draw=red!75,    line width=1.0pt, text width=4.2cm, minimum height=2.0cm},
    inputbox/.style   ={box, fill=green!18,  draw=green!60!black, text width=2.5cm, minimum height=1.2cm},
    outputbox/.style  ={box, fill=red!8,     draw=red!50, text width=2.5cm, minimum height=1.2cm},
    arr/.style        ={->, >=stealth, line width=0.8pt, color=black!75},
    mask/.style       ={->, >=stealth, line width=0.8pt, dashed, color=red!75},
    ptok/.style       ={draw=gray!60, fill=gray!35, minimum width=0.28cm, minimum height=0.28cm, inner sep=0pt, line width=0.4pt},
    targt/.style      ={draw=red!75, fill=red!30,  minimum width=0.28cm, minimum height=0.28cm, inner sep=0pt, line width=0.4pt},
    dimlbl/.style     ={font=\fontsize{9}{10}\selectfont\itshape, color=black!80, inner sep=1pt, fill=white, inner xsep=2pt},
    insetbox/.style   ={draw, rounded corners=3pt, fill=gray!5, draw=gray!50, line width=0.5pt, inner sep=4pt, align=center, font=\footnotesize},
]

\node[inputbox] (input) at (0, 0)
    {\textbf{Bangla Text}\\{\scriptsize UTF-8 string}};

\node[s2] (norm) at (4.6, 0)
    {\textbf{S2: Text Norm.}\\
    {\scriptsize 6-step pipeline}\\
    {\scriptsize digits$\to$num-word$\to$Unicode}\\
    {\scriptsize $\to$honorific$\to$sent.\ seg.}};

\node[s1] (tok) at (9.8, 0)
    {\textbf{S1: Tokenizer Ext.}\\
    {\scriptsize vocab $704{+}1{,}750{+}1{,}786{=}\mathbf{4{,}240}$}\\
    {\scriptsize merge-consistent filter:}\\
    {\scriptsize $(a,b){\to}ab \text{ iff } \{a,b,ab\}{\subseteq}V$}};

\node[insetbox, text width=4.2cm, minimum height=1.0cm, inner sep=2pt] (embmat) at (15.2, 0)
    {\textbf{Embedding resize}\\[-2pt]
    \begin{tikzpicture}[baseline=0pt, sharp corners, every path/.style={rounded corners=0pt}]
        \path[fill=blue!35, draw=blue!75, line width=0.35pt]
            (0, 0.18) -- (0.55, 0.18) -- (0.55, 0.28) -- (0, 0.28) -- cycle;
        \path[fill=orange!35, draw=orange!75, line width=0.35pt]
            (0, -0.12) -- (0.55, -0.12) -- (0.55, 0.18) -- (0, 0.18) -- cycle;
        \node[anchor=west, font=\scriptsize, color=blue!70!black]   at (0.62, 0.23)  {704: copied};
        \node[anchor=west, font=\scriptsize, color=orange!75!black] at (0.62, 0.03) {3{,}536: $\mu$-init};
    \end{tikzpicture}};

\draw[arr] (input.east) -- (norm.west);
\node[dimlbl] at (2.3, 0.45) {raw};
\draw[arr] (norm.east)  -- (tok.west);
\node[dimlbl] at (7.2, 0.95) {\bn{[bn]}+norm};

\draw[->, color=orange!70, line width=0.5pt, >=stealth]
    (tok.east) -- node[dimlbl, text=orange!75!black, midway, above=25pt] {$|V|{=}4{,}240$} (embmat.west);

\node[s3] (t3) at (5.1, -2.8)
    {\textbf{S3: T3 Transformer} {\footnotesize\color{red!75!black}$\nabla$ trainable}\\
    {\scriptsize 520M params, Llama-style AR}\\
    {\scriptsize text-emb + out-head $704{\to}4{,}240$}\\
    {\scriptsize new rows: centroid mean-init}\\
    {\scriptsize loss: $\mathcal{L}_{\text{text}} + \mathcal{L}_{\text{speech}}^{\text{masked}}$}};

\node[frozen] (s3gen) at (10.8, -2.8)
    {\textbf{S3Gen Vocoder} {\footnotesize\color{blue!60!black}$\ast$ frozen}\\
    {\scriptsize 24\,kHz HiFi-GAN}\\
    {\scriptsize $\sim$250M params}};

\node[outputbox] (output) at (15.2, -2.8)
    {\textbf{24\,kHz}\\\textbf{Bangla wav}};

\draw[arr] (t3.east) -- (s3gen.west);
\node[dimlbl] at (8.13, -2.25) {speech tok.};
\node[dimlbl] at (8.0, -3.15) {25\,tok/s};
\draw[arr] (s3gen.east) -- (output.west);
\node[dimlbl] at (13.25, -1.95) {waveform};

\draw[arr] (tok.south) -- ++(0, -0.30) -| (t3.north);

\draw[arr] (embmat.south) -- (embmat.south |- 0,-1.45) -| ([xshift=1.4cm]t3.north);
\node[dimlbl] at (7.45, -1.10) {token IDs $\in[0,4239]$};
\node[dimlbl] at (12.0, -1.45) {used in T3};

\node[inputbox, text width=2.4cm, minimum height=1.6cm] (refaudio) at (0.0, -2.5)
    {\textbf{Reference Audio}\\{\scriptsize 3--10\,s, 16\,kHz}};

\node[frozen, anchor=north, text width=2.9cm, minimum height=1.05cm] (voiceenc) at (-1.4, -4.8)
    {\textbf{Voice Encoder}{\scriptsize\color{blue!60!black}\,$\ast$}\\{\scriptsize 256-d spk.\ emb}};
\node[frozen, anchor=north, text width=2.9cm, minimum height=1.05cm] (s3tok)    at (1.9, -4.8)
    {\textbf{S3 Tokenizer}{\scriptsize\color{blue!60!black}\,$\ast$}\\{\scriptsize 25\,tok/s, $|V_s|{=}8{,}194$}};

\draw[arr] (refaudio.south) -- ++(0,-0.5) -| (voiceenc.north);
\draw[arr] (refaudio.south) -- ++(0,-0.5) -| (s3tok.north);

\draw[->, >=stealth, line width=0.7pt, color=purple!70]
    ([xshift=-0.5cm]voiceenc.north east) -- ++(0, 0.25)
    -| ([xshift=-0.8cm]t3.south);
\node[font=\fontsize{11}{12}\selectfont, color=purple!70, fill=white, inner sep=1pt]
    at (3.10, -4.27) {spk.\ emb};
\draw[->, >=stealth, line width=0.8pt, color=red!75]
    (s3tok.north east) -- ++(0, 0.0)
    node[font=\fontsize{9}{10}\selectfont, midway, right, fill=white, inner sep=1pt] {prompt (masked)}
    -| ([xshift=0.6cm]t3.south);

\node[insetbox, anchor=north, text width=4.5cm] at (10.0, -4.1)
    {\textbf{Prompt-mask + loss}\\[2pt]
    \begin{tikzpicture}[baseline=0pt]
        \foreach \x in {0,1,2,3} {\node[ptok] at (\x*0.28, 0) {};}
        \foreach \x in {0,...,5} {\node[targt] at (1.15 + \x*0.28, 0) {};}
        \draw[decorate, decoration={brace, amplitude=2pt}, line width=0.4pt, color=gray!70]
            (-0.13, -0.19) -- (0.97, -0.19)
            node[midway, below=1pt, font=\scriptsize, color=gray!70] {masked};
        \draw[decorate, decoration={brace, amplitude=2pt}, line width=0.4pt, color=red!70]
            (1.02, -0.19) -- (2.59, -0.19)
            node[midway, below=1pt, font=\scriptsize, color=red!70] {predicted};
    \end{tikzpicture}\\[8pt]
    {\footnotesize $\mathcal{L} = \mathcal{L}_{\text{text}} + \mathcal{L}_{\text{speech}}^{\text{masked}}$}};

\node[s1, text width=0.40cm, minimum height=0.32cm, inner sep=1pt] (LS1) at (-0.5, -7.0) {};
\node[right=0.10cm of LS1, font=\footnotesize] {S1 Tokenizer ext.};

\node[s2, text width=0.40cm, minimum height=0.32cm, inner sep=1pt] (LS2) at (2.9, -7.0) {};
\node[right=0.10cm of LS2, font=\footnotesize] {S2 Text norm.};

\node[s3, text width=0.40cm, minimum height=0.32cm, inner sep=1pt] (LS3) at (5.5, -7.0) {};
\node[right=0.10cm of LS3, font=\footnotesize] {S3 Prompt-masked loss};

\node[frozen, text width=0.40cm, minimum height=0.32cm, inner sep=1pt] (LF) at (10.5, -7.0) {};
\node[right=0.10cm of LF, font=\footnotesize] {Frozen ($\ast$)};

\node[font=\footnotesize, anchor=west, color=red!75!black] at (12.9, -7.0) {$\nabla$ = trainable};

\node[font=\footnotesize, anchor=west, color=purple!70] at (15.8, -7.0) {- - = conditioning};

\end{tikzpicture}%
}
\vspace{-2pt}      
\caption{\sys{} architecture (\S\ref{sec:finetuning}). Only T3 is trainable ($\nabla$); the voice encoder and S3Gen vocoder stay frozen ($\ast$). The first $75$ prompt tokens are masked from $\mathcal{L}_{\text{speech}}^{\text{masked}}$. Inset: $704$ pretrained embedding rows copied, $3{,}536$ new rows mean-initialized from centroid $\mu$.}
\label{fig:architecture}
\vspace{-1.5pt}
\end{figure*}
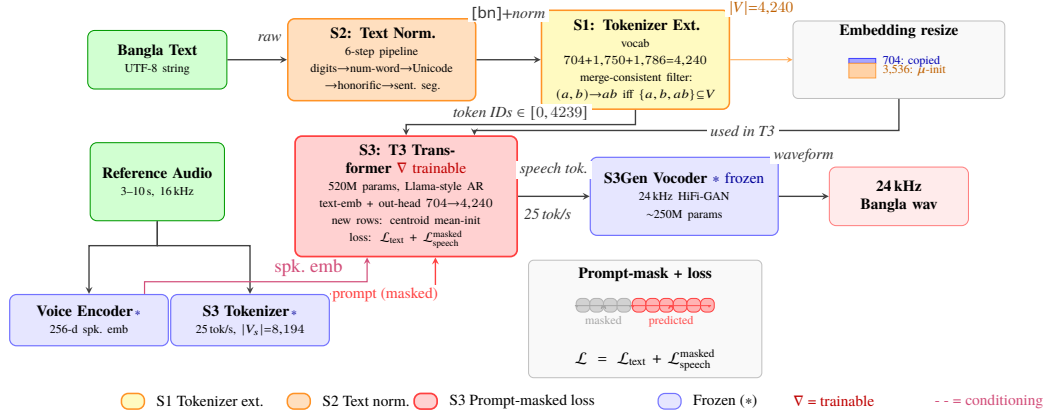

\paragraph{Chatterbox architecture.}
Chatterbox \cite{chatterboxtts2025} comprises (1)~\textbf{T3}, a 520M-parameter Llama-style autoregressive transformer predicting text and speech tokens jointly; (2)~\textbf{S3Gen}, a flow-matching mel decoder with a HiFi-GAN \cite{kong2020hifigan} vocoder at 24\,kHz; and (3)~a \textbf{voice encoder} extracting 256-dim speaker embeddings from reference audio (full architecture in Table~\ref{tab:hyperparams}).
Zero-shot cloning conditions T3 on speaker embeddings plus a short prompt-token slice (3\,s for our Bangla setting; cf.\ stock Chatterbox 6\,s).
Fig.~\ref{fig:architecture} shows our modifications.

\subsection{Fine-tuning pipeline}
\label{sec:finetuning}


Our fine-tuning pipeline has 3 stages: a merge-consistent tokenizer extension, Bangla text normalization, and prompt-masked dual-loss training.

\textbf{Stage 1 -- Merge-consistent Bangla vocabulary extension.} \emph{Why:} the pretrained BPE vocabulary cannot encode Bengali script, and naively imported merge rules surface as UNK errors precisely on conjunct-bearing words (App.~\ref{app:stage1}).
We extend Chatterbox's tokenizer from $704$ to $4{,}240$ tokens by adding $1{,}750$ multilingual base units and $1{,}786$ Bangla units (full vocabulary breakdown in App.~\ref{app:stage1}).
A naive merge import produces invalid rules: a merge $(a,b){\to}ab$ requires $a$, $b$, \emph{and} $ab$ to coexist in the target vocab, so we apply a \emph{merge-consistent filter} that drops merges failing this check, preventing tokenization breakage at inference (alternative extension strategies and their trade-offs: App.~\ref{app:tok_alt}).
The T3 text embedding and output head are resized $704 \to 4{,}240$: the 704 overlapping rows are copied from the pretrained checkpoint; the remaining 3{,}536 rows are \emph{mean-initialized} to the centroid of the 704 pretrained rows, preserving the pretrained embedding geometry rather than injecting random-init variance.

\textbf{Stage 2 --- Bangla text normalization with honorific expansion.} \emph{Why:} unexpanded numerals and honorific abbreviations are verbalized incorrectly at synthesis; removing this stage is the largest CER regression in Table~\ref{tab:ablation}.
The 6-step pipeline covers digit conversion, colon-numeral rewriting, numeral-to-word expansion (\texttt{bnnumerizer} \cite{bnnumerizer}), per-word Unicode normalization (\texttt{bnunicodenormalizer} \cite{bnunicodenormalizer}), a 47-entry dictionary of religious honorific abbreviations absent from open packages (e.g., \bn{সাঃ}$\to$\bn{সাল্লাল্লাহু আলাইহি ওয়া সাল্লাম}), and sentence segmentation; a worked example and the dictionary are in App.~\ref{app:stage2}.

\textbf{Stage 3 -- Prompt-masked dual-loss T3 fine-tuning.}
\emph{Why:} without the prompt mask the autoregressive decoder minimizes loss by copying prompt tokens instead of conditioning on the speaker embedding --- the largest single SECS drop in Table~\ref{tab:ablation} ($0.627{\to}0.51$).
We fine-tune only T3 (520M) while freezing the voice encoder (preserves the pretrained speaker space) and S3Gen (stabilizes the token-to-waveform mapping); ablations (Section~\ref{sec:experiments}) confirm encoder freezing as a safe default and the prompt-masked loss as the decisive component for cloning fidelity.
The training objective is a \emph{prompt-masked} dual cross-entropy loss:
{\setlength{\abovedisplayskip}{4pt}%
\setlength{\belowdisplayskip}{4pt}%
\setlength{\abovedisplayshortskip}{2pt}%
\setlength{\belowdisplayshortskip}{2pt}%
\begin{equation}
\mathcal{L} = \mathcal{L}_{\text{text}} + \mathcal{L}_{\text{speech}}^{\text{masked}},
\label{eq:loss}
\end{equation}
where $\mathcal{L}_{\text{text}}$ is the next-token cross-entropy on the text tokens (T3 predicts text and speech tokens jointly; this term trains the text half), and
\begin{equation}\small
\mathcal{L}_{\text{speech}}^{\text{masked}} = -\!\!\!\! \sum_{t=L_{\text{prompt}}+1}^{T} \!\!\!\! \log p_\theta(s_t \mid s_{<t}, \mathbf{e}_{\text{spk}}, \mathbf{c}_{\text{text}})
\label{eq:loss_speech}
\end{equation}}
is the same cross-entropy on the speech tokens with the prompt positions masked out: $s_t$ denotes the $t$-th speech token and $p_\theta$ the probability the T3 model assigns to the next token.
The first $L_{\text{prompt}}{\approx}75$ speech positions (3\,s at 25\,tok/s) are excluded via \texttt{ignore\_index} alongside right-padding masking; $T$ is the speech-token length, $\mathbf{e}_{\text{spk}}$ the frozen speaker embedding, and $\mathbf{c}_{\text{text}}$ the normalized text tokens; all symbols are listed in Table~\ref{tab:abbrev} (App.~\ref{app:abbrev}).
The same failure mode is addressed in BnTTS \cite{bntts2025} and VALL-E \cite{wang2023valle}.
Because the voice encoder and S3Gen remain frozen, zero-shot cloning is preserved at inference from any 3--10\,s Bangla reference; all 3 stages are independently ablated in \S\ref{sec:experiments}.

\subsection{Long-form Bangla inference}
\label{sec:longform}

Multi-sentence paragraphs exceed our 850-token per-utterance cap ($\approx$34\,s at 25\,tok/s), drift mid-generation, or emit EOS prematurely. Hence, we introduce a \emph{chunked long-form inference pipeline}:
\begin{enumerate}[wide=0pt,leftmargin=0pt,labelindent=0pt,topsep=0pt,itemsep=0pt,parsep=0pt]
\item \textbf{Sentence segmentation and per-utterance synthesis.} The paragraph is split on Bangla terminators (\bn{।}, \texttt{!}, \texttt{?}) and each segment is synthesized with the same speaker embedding and prompt, preserving voice identity across chunks.
\item \textbf{Silero-VAD tail trim.} We trim trailing silence per chunk by Silero VAD \cite{silero_vad}.
\item \textbf{Deterministic inter-sentence pause.} Chunks are joined by a 200\,ms zero-pad pause, rendering the \textit{dari} \bn{।} prosodic reset.
\end{enumerate}

Inference uses Bangla-calibrated hyperparameters (Table~\ref{tab:hyperparams}, inference section): a lower temperature (0.3 vs.\ Chatterbox's English default 0.8) counters English-phoneme hallucination on unfamiliar conjuncts, and a raised minimum-new-tokens floor (150) counters premature EOS on short references. A dedicated chunked-vs-single-pass ablation is reported in \S\ref{sec:analysis}.

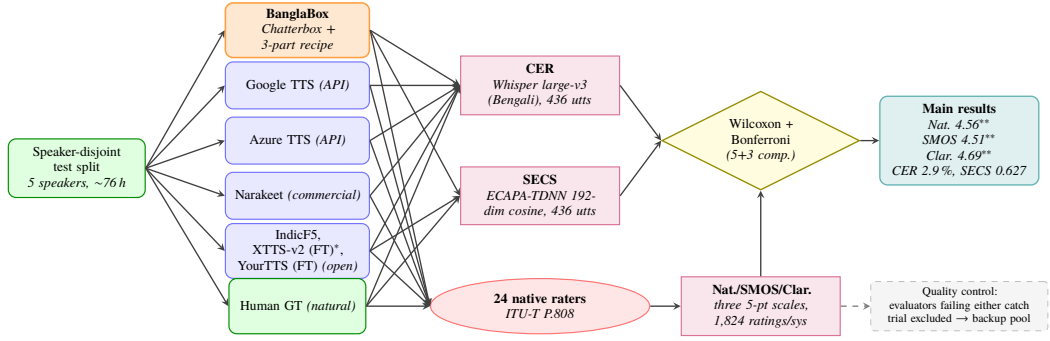
\begin{figure*}[!t]
\centering
\resizebox{0.86\textwidth}{!}{%
\begin{tikzpicture}[
    every node/.style={font=\scriptsize, align=center},
    src/.style       ={draw, rounded corners=4pt, fill=green!12,  draw=green!55!black, line width=0.6pt, text width=2.2cm, minimum height=1.0cm, inner sep=4pt},
    sys/.style       ={draw, rounded corners=4pt, fill=blue!9,    draw=blue!55,        line width=0.6pt, text width=2.4cm, minimum height=0.85cm, inner sep=3pt},
    ours/.style      ={draw, rounded corners=4pt, fill=orange!18, draw=orange!75,      line width=0.8pt, text width=2.4cm, minimum height=0.85cm, inner sep=3pt},
    metric/.style    ={draw, fill=purple!10, draw=purple!55,      line width=0.6pt, text width=2.6cm, minimum height=1.0cm, inner sep=4pt},
    rater/.style     ={draw, ellipse, fill=red!10,  draw=red!55,  line width=0.6pt, text width=2.6cm, inner sep=3pt, minimum height=1.0cm},
    stat/.style      ={draw, diamond, aspect=2, fill=yellow!15, draw=olive, line width=0.7pt, inner sep=2pt, minimum height=1.0cm, text width=1.6cm},
    outbox/.style    ={draw, rounded corners=4pt, fill=teal!12, draw=teal!60, line width=0.7pt, text width=2.6cm, minimum height=1.6cm, inner sep=4pt},
    arr/.style       ={->, >=stealth, line width=0.7pt, color=black!75},
    sidearr/.style   ={->, >=stealth, line width=0.7pt, color=black!50, dashed},
]

\node[src] (testset) at (0, -1)
    {Speaker-disjoint test split \\ \textit{5 speakers, $\sim$76\,h}};

\node[ours] (banglabox) at (4.0,  1.5)  {\textbf{\sys} \\ \textit{Chatterbox + 3-part recipe}};
\node[sys]  (google)    at (4.0,  0.5)  {Google TTS \textit{(API)}};
\node[sys]  (azure)     at (4.0, -0.5)  {Azure TTS \textit{(API)}};
\node[sys]  (narakeet)  at (4.0, -1.5)  {Narakeet \textit{(commercial)}};
\node[sys]  (indicf5)   at (4.0, -2.5)  {IndicF5, \\ \mbox{XTTS-v2} (FT)$^{\ast}$, \\ YourTTS (FT) \textit{(open)}};
\node[src]  (human)     at (4.0, -3.5)  {Human GT \textit{(natural)}};

\draw[arr] (testset.east) -- (banglabox.west);
\draw[arr] (testset.east) -- (google.west);
\draw[arr] (testset.east) -- (azure.west);
\draw[arr] (testset.east) -- (narakeet.west);
\draw[arr] (testset.east) -- (indicf5.west);
\draw[arr] (testset.east) -- (human.west);

\node[metric] (cer)    at (8.4,  0.5)  {\textbf{CER} \\ \textit{Whisper large-v3 (Bengali), 436 utts}};
\node[metric] (secs)   at (8.4, -1.5)  {\textbf{SECS} \\ \textit{ECAPA-TDNN 192-dim cosine, 436 utts}};

\draw[arr] (banglabox.east) -- (cer.west);
\draw[arr] (google.east)    -- (cer.west);
\draw[arr] (azure.east)     -- (cer.west);
\draw[arr] (narakeet.east)  -- (cer.west);
\draw[arr] (indicf5.east)   -- (cer.west);
\draw[arr] (human.east)     -- (cer.west);
\draw[arr] (banglabox.east) -- (secs.west);
\draw[arr] (indicf5.east)   -- (secs.west);
\draw[arr] (human.east)     -- (secs.west);

\node[rater] (raters)  at (8.4, -3.5)
    {\textbf{24 native raters} \\ \textit{ITU-T P.808}};

\node[metric] (subj) at (12.4, -3.5)
    {\textbf{Nat./SMOS/Clar.} \\ \textit{three 5-pt scales, 1{,}824 ratings/sys}};

\draw[arr] (banglabox.east) -- (raters.west);
\draw[arr] (google.east)    -- (raters.west);
\draw[arr] (azure.east)     -- (raters.west);
\draw[arr] (narakeet.east)  -- (raters.west);
\draw[arr] (indicf5.east)   -- (raters.west);
\draw[arr] (human.east)     -- (raters.west);
\draw[arr] (raters.east)    -- (subj.west);

\node[stat] (stat) at (12.4, -0.5)
    {Wilcoxon + Bonferroni \\ \textit{(5+3 comp.)}};

\draw[arr] (cer.east)  -- (stat.west);
\draw[arr] (secs.east) -- (stat.west);
\draw[arr] (subj.north) -- (stat.south);

\node[outbox] (results) at (16.0, -0.5)
    {\textbf{Main results} \\ \textit{Nat.\ 4.56$^{**}$} \\ \textit{SMOS\ 4.51$^{**}$} \\ \textit{Clar.\ 4.69$^{**}$} \\ \textit{CER 2.9\,\%, SECS 0.627}};

\draw[arr] (stat.east) -- (results.west);

\node[draw, dashed, rounded corners=2pt, fill=gray!8, draw=gray!50, inner sep=3pt, font=\tiny, text width=3.0cm]
    (qc) at (16.0, -3.5)
    {Quality control: \\ evaluators failing either catch trial excluded $\to$ backup pool};

\draw[sidearr] (subj.east) -- (qc.west);

\end{tikzpicture}%
}
\caption{BanglaEval evaluation pipeline (\S\ref{sec:experiments}). $^{\ast}$XTTS-v2 (FT) is scored on objective and cloning metrics only. Character error rate (CER); speaker-encoder cosine similarity (SECS); speaker-similarity mean opinion score (SMOS). Results in Table~\ref{tab:main_results}.}
\label{fig:experiment_pipeline}
\end{figure*}

\section{Evaluation Protocol -- BanglaEval}
\new{\noindent In addition to the protocol below, \S\ref{sec:stress} reports three evaluations added in this revision: a stress test over difficult real-world text, cross-validation on the public BnTTS benchmarks, and a register control on naturally occurring Bangla.}

\label{sec:experiments}

Our evaluation protocol, \emph{BanglaEval}, follows best practices in TTS evaluation \cite{kirkland2023mos} and the recent position paper of \citet{yang2025responsibleeval} on responsible evaluation for text-to-speech in the 2025--2026 era of foundation models.

\begin{table}[!t]
\centering
\footnotesize
\setlength{\tabcolsep}{3pt}
\begin{tabular*}{\columnwidth}{@{\extracolsep{\fill}}lccc@{}}
\toprule
\textbf{System} & \textbf{Nat.}$\uparrow$ & \textbf{SMOS}$\uparrow$ & \textbf{Clar.}$\uparrow$ \\
\midrule
Google TTS       & 4.27{\scriptsize$\pm$0.17} & 4.17{\scriptsize$\pm$0.13} & 4.51{\scriptsize$\pm$0.13} \\
Azure TTS        & 4.13{\scriptsize$\pm$0.14} & 4.07{\scriptsize$\pm$0.11} & 4.44{\scriptsize$\pm$0.15} \\
Narakeet         & 3.97{\scriptsize$\pm$0.16} & 3.71{\scriptsize$\pm$0.13} & 4.02{\scriptsize$\pm$0.12} \\
IndicF5          & 3.89{\scriptsize$\pm$0.15} & 3.55{\scriptsize$\pm$0.14} & 3.95{\scriptsize$\pm$0.12} \\
YourTTS (FT)     & 4.18{\scriptsize$\pm$0.14} & 3.08{\scriptsize$\pm$0.15} & 4.39{\scriptsize$\pm$0.13} \\
\new{XTTS-v2 (FT)}     & \new{4.21} & \new{---} & \new{4.35} \\
\midrule
\sys             & \textbf{4.56}{\scriptsize$\pm$0.16}$^{**}$ & \textbf{4.51}{\scriptsize$\pm$0.10}$^{**}$ & \textbf{4.69}{\scriptsize$\pm$0.08}$^{**}$ \\
\midrule
Human\textsuperscript{\dag} & 4.76{\scriptsize$\pm$0.15} & 4.79{\scriptsize$\pm$0.14} & 4.84{\scriptsize$\pm$0.05} \\
\bottomrule
\end{tabular*}
\caption{\textbf{Subjective} results from 24 raters (76 utts., 1{,}824 ratings/system). \textbf{Bold}: best synthesized; \textsuperscript{\dag}human upper bound. $^{**}p{<}0.01$ vs.\ baselines (Wilcoxon + Bonferroni). CI computation, Krippendorff's $\alpha$ and exact statistics: App.~\ref{app:stats}. \new{The XTTS-v2 (FT) row is new in this revision.}}
\label{tab:main_results}
\end{table}

\begin{table}[!t]
\centering
\footnotesize
\setlength{\tabcolsep}{4pt}
\begin{tabular*}{\columnwidth}{@{\extracolsep{\fill}}lccc@{}}
\toprule
\textbf{System} & \textbf{NISQA}$\uparrow$ & \textbf{CER}$\downarrow$ & \textbf{SECS}$\uparrow$ \\
\midrule
Google TTS       & 4.52 & 2.2 & --- \\
Azure TTS        & 4.43 & 2.7 & --- \\
Narakeet         & 4.37 & 3.3 & --- \\
IndicF5          & 4.13 & 3.8 & 0.37 \\
YourTTS (FT)     & 3.69 & 3.1 & 0.44 \\
XTTS-v2 (FT)     & 4.02 & 3.6 & 0.47 \\
\midrule
\sys             & \textbf{4.59} & 2.9 & \textbf{0.627}$^{**}$ \\
\midrule
Human\textsuperscript{\dag} & 4.81 & 2.1 & 0.85 \\
\bottomrule
\end{tabular*}
\caption{\textbf{Objective} results: NISQA \cite{mittag2021nisqa} (reference-free MOS), CER (Whisper large-v3, Bengali), SECS (ECAPA-TDNN, 192-dim) over 436 utts. SECS n/a for commercial (no speaker adaptation). Markers: $^{**}p{<}0.01$ vs.\ each cloning-capable baseline (Wilcoxon signed-rank on per-utterance SECS, Bonferroni over three comparisons; App.~\ref{app:stats}).}
\label{tab:main_results_obj}
\end{table}

\begin{table}[t]
\centering
\footnotesize
\setlength{\tabcolsep}{4pt}
\begin{tabular*}{\columnwidth}{@{\extracolsep{\fill}}llcc@{}}
\toprule
\textbf{System} & \textbf{Ref.} & \textbf{SECS}$\uparrow$ & \textbf{SMOS}$\uparrow$ \\
\midrule
\multirow{2}{*}{IndicF5}   & 5\,s  & 0.31 & 3.46{\scriptsize$\pm$0.14} \\
                            & 10\,s & 0.37 & 3.55{\scriptsize$\pm$0.14} \\
\midrule
\multirow{2}{*}{XTTS-v2 (FT)}   & 5\,s  & 0.41 & 3.88{\scriptsize$\pm$0.15} \\
                            & 10\,s & 0.47 & 3.79{\scriptsize$\pm$0.13} \\
\midrule
\multirow{2}{*}{YourTTS (FT)}   & 5\,s  & 0.42 & 3.01{\scriptsize$\pm$0.15} \\
                            & 10\,s & 0.44 & 3.08{\scriptsize$\pm$0.15} \\
\midrule
\multirow{2}{*}{\sys}      & 5\,s  & 0.58$^{**}$ & 4.47$^{**}${\scriptsize$\pm$0.10} \\
                            & 10\,s & \textbf{0.627}$^{**}$ & \textbf{4.51}$^{**}${\scriptsize$\pm$0.10} \\
\bottomrule
\end{tabular*}
\caption{Zero-shot voice cloning on held-out speakers, 5\,s vs.\ 10\,s reference. \textbf{Bold}: best overall. Markers: $^{*}p{<}0.05$, $^{**}p{<}0.01$ vs.\ each cloning-capable baseline (Wilcoxon + Bonferroni). Metric defs in \S\ref{sec:experiments}.}
\label{tab:cloning}
\end{table}

\subsection{Training setup}
We fine-tune \sys for 30--50 epochs on the speaker-disjoint train split (Table~\ref{tab:dataset_stats}) using the configuration in Table~\ref{tab:hyperparams}. Inference uses the Bangla-calibrated hyperparameters and chunked pipeline of \S\ref{sec:longform} across all conditions.

\subsection{Metrics and rater protocol}

\textbf{Recruitment.} 24 native Bangla speakers (14\,M/10\,F, ages 22--45, all from Bangladesh), screened for L1 Bangla, a 5-utterance listening pre-screen ($\geq$80\% accuracy) and no reported hearing impairment (App.~\ref{app:raters}).
\textbf{Listening environment.} Following ITU-T P.808 \cite{itu_p808}: closed-back headphones, verified quiet environment, all stimuli RMS-normalized to $-23$\,LUFS.
\textbf{Stimuli.} Each evaluator rates 76 utterances drawn from a balanced subjective evaluation set; each utterance is rated by all 24 evaluators across systems, yielding $76\times24=1{,}824$ ratings per system on 3 independent 5-point Likert scales:
\textbf{Naturalness} (resemblance to human speech), \textbf{Clarity} (intelligibility and articulation) and \textbf{SMOS} (speaker similarity, rated against a reference).
\textbf{Quality control.} Two catch trials per session; evaluators failing either are excluded and replaced (App.~\ref{app:catch}). \textbf{Statistics.} 95\% CIs and Wilcoxon signed-rank with Bonferroni correction (5 comparisons for Nat., SMOS, Clar.; 3 for SECS); ground truth included as an upper bound.
For objective evaluation we report \textbf{CER} via Whisper large-v3 \cite{whisper2023} with language forced to Bengali over all 436 test utterances (ground-truth CER confirms ASR reliability), \textbf{NISQA} \cite{mittag2021nisqa} as a reference-free deep neural MOS-scale quality estimator, and \textbf{SECS} via ECAPA-TDNN \cite{desplanques2020ecapa} (\texttt{speechbrain/spkrec-ecapa-voxceleb}, 192-dim).\footnote{BnTTS reports SECS with ECAPA2 \cite{thienpondt2023ecapa2} and a different backbone (XTTS vs.\ Chatterbox), so the $0.548$ few-shot figure we cite is indicative, not head-to-head.}
Fig.~\ref{fig:experiment_pipeline} summarizes the pipeline.

\subsection{Baselines}

\textbf{Google TTS} (Cloud API, Bangla); \textbf{Azure TTS} (Microsoft Cognitive Services, Bangla); \textbf{Narakeet}~\cite{narakeet} (commercial neural); \textbf{IndicF5} \cite{indicf5} (AI4Bharat zero-shot Indic TTS); \textbf{YourTTS} \cite{casanova2022yourtts} fine-tuned on our corpus; \textbf{XTTS-v2} \cite{casanova2024xtts} fine-tuned on the identical corpus (BnTTS's backbone family --- the strictest comparable-data setting; objective and cloning only); and \textbf{Human} natural recordings (upper bound).
Our baseline set includes the same commercial systems as BnTTS \cite{bntts2025} plus four additional comparisons; direct BnTTS comparison is precluded by unreleased weights.

\subsection{Main results}

Tables~\ref{tab:main_results} and~\ref{tab:main_results_obj} report subjective metrics over 76 domain-balanced utterances and objective metrics over a 436-utterance domain-stratified subset of the speaker-disjoint test split.
\sys achieves the highest Naturalness, SMOS and Clarity; every pairwise difference is significant, as are the SECS differences against each cloning-capable baseline ($p{<}0.01$). CER is competitive with Google and Azure and ahead of all open baselines, and NISQA \cite{mittag2021nisqa} also ranks \sys{} highest.
Table~\ref{tab:pairwise} gives exact $p$-values and effect sizes; all 15 comparisons clear $\alpha/5{=}0.01$. By domain (Table~\ref{tab:perdomain}), Naturalness stays within a narrow band across all 7, so no domain is left behind by the balanced corpus.

\subsection{Voice cloning}

Table~\ref{tab:cloning} reports zero-shot voice-cloning results.
With a 10-second reference from held-out speakers, \sys substantially surpasses YourTTS-FT, IndicF5 and the same-corpus XTTS-v2 fine-tune; the 5\,s setting preserves \sys{}'s lead.
We further report the indicative comparison \sys $0.627$ (ECAPA-TDNN) vs.\ BnTTS's reported few-shot headline $0.548$ (ECAPA2; encoders differ), comparable to prior few-shot results, trained on $520.3$\,h vs.\ $3.85$k\,h (Table~\ref{tab:dataset_compare}).

\begin{figure}[t]
\centering
\begin{tikzpicture}
\definecolor{clrMOS}{RGB}{83,74,183}
\definecolor{clrCER}{RGB}{216,90,48}
\begin{axis}[
    width=0.86\columnwidth,
    height=5.0cm,
    xlabel={Corpus size (hours)},
    ylabel={\textcolor{clrMOS}{Naturalness ($\uparrow$)}},
    xlabel style={font=\normalsize},
    ylabel style={font=\normalsize},
    xticklabel style={font=\normalsize},
    yticklabel style={font=\normalsize, color=clrMOS},
    axis y line*=left,
    ymin=4.0, ymax=4.7,
    xmin=50, xmax=570,
    xtick={100,200,520},
    grid=major,
    grid style={gray!15},
    legend style={font=\footnotesize, at={(0.02,0.98)}, anchor=north west, fill=white, fill opacity=0.9, draw=gray!30},
]
\addplot[name path=mos, solid, very thick, color=clrMOS, mark=o, mark size=3pt, mark options={fill=clrMOS}] coordinates {
    (100, 4.18) (200, 4.36) (520, 4.56)
};
\addlegendentry{\textcolor{clrMOS}{Naturalness}}
\addplot[only marks, mark=star, mark size=5pt, fill=clrMOS, color=clrMOS] coordinates {(520, 4.56)};
\addplot[name path=bottom, draw=none] coordinates {(100,4.0) (200,4.0) (520,4.0)};
\addplot[clrMOS, fill opacity=0.08] fill between[of=mos and bottom];
\end{axis}
\begin{axis}[
    width=0.86\columnwidth,
    height=5.0cm,
    ylabel={\textcolor{clrCER}{CER (\%) ($\downarrow$)}},
    ylabel style={font=\normalsize},
    yticklabel style={font=\normalsize, color=clrCER},
    axis y line*=right,
    axis x line=none,
    ymin=2, ymax=5,
    xmin=50, xmax=570,
    legend style={font=\footnotesize, at={(0.98,0.98)}, anchor=north east, fill=white, fill opacity=0.9, draw=gray!30},
]
\addplot[solid, very thick, color=clrCER, mark=square*, mark size=3pt, mark options={fill=clrCER}] coordinates {
    (100, 4.3) (200, 3.4) (520, 2.9)
};
\addlegendentry{\textcolor{clrCER}{CER}}
\addplot[only marks, mark=star, mark size=5pt, fill=clrCER, color=clrCER] coordinates {(520, 2.9)};
\end{axis}
\end{tikzpicture}
\caption{Data scaling on 100/200/520\,h subsets. Naturalness (purple, left) keeps improving; character error rate (CER, orange, right) saturates beyond 200\,h. Stars: full 520\,h configuration.}
\label{fig:scaling}
\end{figure}
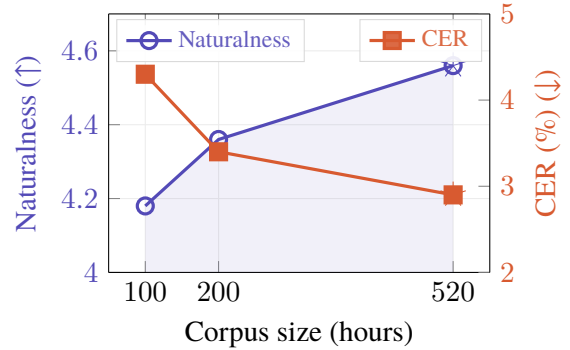

\begin{table}[t]
\centering
\footnotesize
\setlength{\tabcolsep}{3pt}
\begin{tabular*}{\columnwidth}{@{\extracolsep{\fill}}lccc@{}}
\toprule
\textbf{Configuration} & \textbf{Nat.}$\uparrow$ & \textbf{CER}$\downarrow$ & \textbf{SECS}$\uparrow$ \\
\midrule
Full \sys              & \textbf{4.56}{\scriptsize$\pm$0.16} & \textbf{2.9} & \textbf{0.627} \\
\midrule
$-$ Tokenizer ext.\    & 4.39$^{**}${\scriptsize$\pm$0.17} & 3.3 & 0.60$^{*}$ \\
$-$ Text normalization & 4.23$^{***}${\scriptsize$\pm$0.17} & 4.4 & 0.56$^{**}$ \\
$-$ Prompt mask        & 4.34$^{**}${\scriptsize$\pm$0.16} & 3.9 & 0.51$^{**}$ \\
$+$ Trainable enc.\    & 4.55{\scriptsize$\pm$0.15} & 3.0 & 0.62 \\
\midrule
100\,h subset          & 4.18{\scriptsize$\pm$0.18} & 4.3 & 0.54 \\
200\,h subset          & 4.36{\scriptsize$\pm$0.16} & 3.4 & 0.59 \\
520\,h (full)          & \textbf{4.56}{\scriptsize$\pm$0.16} & \textbf{2.9} & \textbf{0.627} \\
\bottomrule
\end{tabular*}
\caption{Ablation study. Upper: each row removes one component; ``$+$ Trainable enc.''\ un-freezes the voice encoder and ``$-$ Tokenizer ext.''\ reverts to a naive extension. Lower: data-scaling subsets. \textbf{Bold}: full \sys. Markers: $^{*}p{<}0.05$, $^{**}p{<}0.01$, $^{***}p{<}0.001$ (Wilcoxon + Bonferroni).}
\label{tab:ablation}
\end{table}

\subsection{Ablations}
Table~\ref{tab:ablation} reports component-level ablations and data-scaling subsets.
Removing text normalization causes the largest CER and Naturalness drop ($p{<}0.001$), so honorific and numeral expansion matters most for intelligibility; removing the prompt mask causes the largest SECS drop, confirming it as the key ingredient for cloning fidelity; and reverting Stage~1 to a naive extension degrades all metrics. Un-freezing the voice encoder changes Nat./SECS by $-0.01$. \new{We previously described this as ``within noise'' without having measured run-to-run variation; we now report only the observed change, and the three-seed comparison below supplies a direct estimate of that variation.} Scaling from 100\,h to 520\,h improves all metrics with distinct patterns: CER and SECS show diminishing returns beyond 200\,h while Naturalness keeps gaining, supporting our curation-first thesis.

\section{Analysis}
\new{\noindent\textbf{Inter-rater reliability.} Krippendorff's $\alpha$ (ordinal, 24 raters) is
$0.44$ for Naturalness, $0.50$ for SMOS and $0.44$ for Clarity --- the
$0.40$--$0.70$ band typical of MOS studies. This also resolves an apparent
tension in Table~\ref{tab:main_results}: the intervals are \emph{marginal}, so
they carry rater and item variation and are wide, whereas Wilcoxon is
\emph{paired}, cancelling rater severity and item difficulty. A system can
therefore win on nearly every item, and be significant, while the marginal
intervals still overlap.}

\label{sec:analysis}

\textbf{Data scaling.} Fig.~\ref{fig:scaling} plots metrics on 100/200/520\,h subsets, supporting our phonetic-balance thesis: CER saturates while Naturalness persists, contrasting BnTTS's large 3.85k\,h regime on Bangla \cite{bntts2025}.

\textbf{Long-form robustness.}
Above the $850$-token cap single-pass synthesis is unreliable, so the chunked pipeline of \S\ref{sec:longform} is the only workable option; below the cap the two are directly comparable. Over $40$ multi-sentence paragraphs (Table~\ref{tab:longform_abl}) chunking reduces premature truncation and length-driven degradation, at a cost in throughput.

\textbf{Prosody.}
Limitations flag English-like prosody in long utterances, so we measure it directly. Table~\ref{tab:prosody} reports F0 RMSE and duration deviation against the held-out speaker's ground truth, split at the median reference duration. Both degrade with length --- duration deviation doubles --- consistent with timing drifting further from the reference as utterances lengthen.

\begin{table}[t]
\centering
\footnotesize
\setlength{\tabcolsep}{4pt}
\color{v3red}
\resizebox{\columnwidth}{!}{%
\begin{tabular}{@{}lccccc@{}}
\toprule
\textbf{$100$\,h subset} & \textbf{Conj.}$\uparrow$ & \textbf{Rare}$\uparrow$ & \textbf{Nat.}$\uparrow$ & \textbf{CER}$\downarrow$ & \textbf{SECS}$\uparrow$ \\
\midrule
JSD-balanced & \textbf{97.0\%} & \textbf{91.7\%} & 4.18{\scriptsize$\pm$0.18} & 4.3 & 0.54 \\
Random       & 93.5\% & 79.8\% & 4.21{\scriptsize$\pm$0.13} & 7.1 & 0.52 \\
\midrule
\new{JSD-balanced, 3 seeds} & \new{---} & \new{---} & \new{\textbf{4.31}} & \new{---} & \new{---} \\
\new{Random, 3 seeds}       & \new{---} & \new{---} & \new{4.18} & \new{---} & \new{---} \\
\bottomrule
\end{tabular}%
}
\caption{Balanced vs.\ random selection at matched $100$\,h from the same Tier~A pool, trained identically. Coverage: share of the conjunct inventory covered, in full and over its rare tail (App.~\ref{app:balrand}). \new{Seed rows (1234, 2025, 31337) are new in this revision.}}
\label{tab:balrand}
\end{table}

\textbf{Failure modes.}
Three error categories persist (rare three-consonant conjuncts, code-switching boundaries, onomatopoeia); see Appendix~\ref{app:failures}.

\subsection{\new{Robustness, external validation and register control}}
\label{sec:stress}
\label{sec:external}
\label{sec:register}
\new{Aggregate error hides where a system breaks. We built a held-out stress set in seven categories, synthesising each sentence three times without selecting the best. Error stays low across all seven and speaker similarity is stable (Table~\ref{tab:stress}); addresses and spelling errors are hardest, both close to the in-domain reference. The set and its audio are released.}

\begin{table}[t]
\centering
\footnotesize
\color{v3red}
\begin{tabular}{@{}lcc@{}}
\toprule
\textbf{Category} & \textbf{CER (\%)}$\downarrow$ & \textbf{SECS}$\uparrow$ \\
\midrule
Out-of-vocabulary words & 2.2 & .72 \\
Spelling errors         & 2.4 & .72 \\
Proper names            & 1.9 & .72 \\
Code-mixed English      & 1.3 & .71 \\
Complex juktakkhor      & 0.7 & .73 \\
Numerics                & 1.0 & .72 \\
Addresses               & 2.9 & .70 \\
\midrule
In-domain reference     & 0.5 & .70 \\
\bottomrule
\end{tabular}
\caption{\new{Stress test over seven categories of difficult real-world text.}}
\label{tab:stress}
\end{table}

\noindent\new{Our own split cannot establish comparability, so we ran the public BnTTS evaluation sets under our objective protocol: our CER improves over the reported BnTTS figures on all three (Table~\ref{tab:external}); we sampled each set rather than the full version, so rows differ in size. Tier~A scripts are also LLM-generated, so we checked the register separately: on held-out Tier~B text that no model wrote, CER is $2.5\%$ against $2.2\%$ (App., Table~\ref{tab:register}). All outputs are released.}

\section{Conclusion}
\label{sec:conclusion}

We present \sys, a recipe for adapting English-pretrained TTS to an underrepresented language: tokenizer extension, text normalization and prompt-masked fine-tuning over a 520.3\,h JSD-balanced corpus. It beats 5 baselines on all 3 perceptual scales ($p{<}0.01$) and matches prior few-shot speaker similarity \cite{bntts2025} on ${\sim}1/7$ of the audio --- curation, not volume, as the matched-hours experiment isolates. \new{Beyond our own split, it holds on a seven-category stress set, the public BnTTS benchmarks and naturally occurring Bangla; all artifacts are public.}

\section*{Limitations}

\textbf{Modest speaker count.} Our 36-speaker corpus is small relative to VCTK (110) and LibriTTS (2{,}456); broader speaker diversity would likely improve cloning generalization.

\textbf{Tier heterogeneity.} Tier A and Tier B differ in recording conditions and provenance; Tier B was empirically curated (16 of an initial 39 speakers were dropped after degrading downstream metrics), but residual tier variance may persist and should be evaluated per-tier.

\textbf{LLM-generated register.} All Tier A scripts are LLM-generated. Despite a Bangla LM-perplexity gate, a language-ID gate, and a 100\% two-pass human review including a native Bangla linguist, the scripts may still carry translationese or LLM-register phrasing that naturalistic spoken Bangla would not, and this register could propagate to the synthesized speech; Tier B (153.0\,h) is the only non-LLM text in training. The blind evaluation gives indirect evidence that this does not dominate perceived quality --- 24 native raters placed the system within $0.20$ MOS of hidden human recordings (Table~\ref{tab:main_results}) --- but a direct text-level register analysis, comparing word choice, sentence shape and loanword rate against natural Bangla corpora, remains future work. \new{As a first control we evaluated on held-out natural Tier~B text that no language model wrote (\S\ref{sec:register}); error is essentially unchanged, which bounds how far the LLM register can be inflating our results.}

\textbf{No head-to-head BnTTS comparison.} BnTTS weights are not public; direct comparison is precluded by unreleased weights, differing speaker encoders (ECAPA2 vs.\ ECAPA-TDNN), differing model families (XTTS vs.\ Chatterbox), and differing evaluation protocols (\S\ref{sec:experiments}). All BnTTS comparisons in this paper are therefore indicative context, and the hours ratio ($520.3$\,h vs.\ $3.85$k\,h) is reported as a factual property of the two training setups, not as a claim of superiority. \new{We have since narrowed this gap by running BnTTS's own public evaluation sets under our objective protocol (\S\ref{sec:external}), so that comparison no longer rests on figures reported elsewhere. The weights remain unavailable, so a system-internal, same-encoder comparison is still not possible.}

\textbf{Pretraining bias.} \sys occasionally produces English-like prosody in long utterances. \new{We now quantify this rather than only noting it: pitch range, pause length and speaking rate are measured against the held-out human recordings, and the gap grows with utterance length (\S\ref{sec:analysis}).}

\textbf{Text normalization gaps.} Domain-specific abbreviations (medical, legal) and Banglish transliteration are not yet handled.

\textbf{No streaming.} RTF $0.42$ on a single H100 GPU, measured on single-utterance unbatched inference without the chunked pipeline, is sufficient for offline use only. \new{The $0.66$ reported in Table~\ref{tab:longform_abl} is a different measurement --- one pass over 40 multi-sentence paragraphs --- and both were previously labelled ``single-pass''; each figure now states the condition it was measured under.}

\textbf{Evaluation scope.} \sys targets Standard Bangladeshi Bangla (Dhaka register); all evaluators and corpus speakers are from Bangladesh. Generalization to regional varieties spoken in West Bengal, Assam, and Tripura is untested. Our blind test includes a hidden human recording per system, but a MUSHRA-style test --- with a fixed reference and anchor clips --- is a stronger design and the clear next evaluation step.

\textbf{Speaker similarity model.} No Bangla-specific speaker verification model exists; we use language-agnostic ECAPA-TDNN embeddings following standard practice \cite{desplanques2020ecapa}.

Future work includes multi-speaker expansion, code-switching support, light LoRA/adapter-based parameter-efficient tuning \cite{hu2022lora} of the voice encoder and S3Gen, and streaming inference.

\section*{Ethics statement}
\label{sec:ethics}

\textbf{Corpus speaker consent.} All speakers in our corpus provided written informed consent for their recordings to be used in TTS research.

\textbf{Evaluator consent.} The 24 human evaluators in our subjective evaluation study provided informed consent (full protocol in \S\ref{sec:experiments}). The evaluation protocol was reviewed and approved by our institutional ethics board. No personally identifiable information was collected from evaluators.

\textbf{Compensation.} The 24 evaluators were paid USD\,15/hour, exceeding local prevailing-wage thresholds and crowd-work-guideline pay floors for the region; corpus speakers were compensated above local market rates in BDT.

\textbf{Voice cloning risks and technical safeguards.} Zero-shot voice cloning introduces risks of impersonation and non-consensual synthetic speech.
We deploy three concrete safeguards rather than a policy disclaimer.
(i)~\emph{Neural audio watermarking.} Our backbone ships with Resemble AI's Perth implicit watermarker, which imprints an imperceptible, detectable mark on generated audio; our fine-tune inherits it, and the watermark is applied unconditionally to every generated waveform, enabling third-party detection of \sys{} output.
(ii)~\emph{Consent-verified voice sources.} Every Tier~A speaker signed written informed consent before any recording session (App.~\ref{app:qa}); no voice in the corpus was captured without documented consent.
(iii)~\emph{Research-only use terms.} Those consent terms cover research use only and exclude any commercial use of the recordings.

\textbf{Broader impact.} High-quality Bangla TTS enables accessibility (screen readers for visually impaired users), education (audiobook generation), and information access (news synthesis). We believe these benefits substantially outweigh misuse risks when the system is used responsibly.

\section*{Acknowledgments}
Generative LLMs (GPT-4o-mini, GPT-4o, and Claude Opus) were used as components
of the Tier~A script-generation pipeline and for language polish during
manuscript preparation. The authors verified all technical content,
experimental results, and scientific claims.


\clearpage
\appendix
\setlength{\textfloatsep}{12pt plus 2pt minus 2pt}%
\setlength{\intextsep}{12pt plus 2pt minus 2pt}%
\setlength{\floatsep}{12pt plus 2pt minus 2pt}%
\setlength{\dbltextfloatsep}{14pt plus 2pt minus 2pt}%
\setlength{\dblfloatsep}{12pt plus 2pt minus 2pt}%

\section{Implementation Notes}

This section elaborates on the three fine-tuning stages introduced in \S\ref{sec:finetuning}. Table~\ref{tab:hyperparams} lists the complete fine-tuning and inference configuration.

\begin{table}[t]
\centering
\footnotesize
\setlength{\tabcolsep}{3pt}
\begin{tabular}{@{}p{0.38\columnwidth}p{0.56\columnwidth}@{}}
\toprule
\textbf{Parameter} & \textbf{Value} \\
\midrule
Base model           & Chatterbox TTS (T3) \\
Architecture         & Llama-style; 1024-dim, 30 layers, 16 heads \\
Trainable / frozen   & T3 (520M) / voice enc.\ + S3Gen \\
Optimizer            & AdamW ($\beta_1{=}0.9$, $\beta_2{=}0.999$) \\
Peak LR              & $5{\times}10^{-6}$, cosine, 5\% warmup \\
Weight decay         & $1{\times}10^{-2}$ \\
Epochs / batch       & 30--50 / 24 \\
Precision / dropout  & BF16 / 0 \\
Loss                 & $\mathcal{L}_{\text{text}}+\mathcal{L}_{\text{speech}}^{\text{masked}}$ (CE) \\
Text / speech vocab  & 4{,}240 ($1{,}786$ Bangla) / 8{,}194 \\
Speaker embed.\ dim  & 256 (voice enc.) \\
Max text / speech tok & 256 / 850 \\
Prompt duration      & 3.0\,s \\
Sample rate          & 16\,kHz in, 24\,kHz out \\
GPU / hours          & 1$\times$ H100 80\,GB / $\sim$100 H100-h (Chatterbox runs; App.~\ref{app:compute}) \\
\midrule
\multicolumn{2}{@{}l}{\textit{Inference (Bangla-calibrated)}} \\
Temperature          & 0.3 (default 0.8) \\
Min.\ new speech tok & 150 (default 0) \\
CFG / rep.\ penalty  & 0.5 / 1.2 \\
Inter-chunk pause    & 200\,ms \\
\bottomrule
\end{tabular}
\caption{Fine-tuning and inference configuration for \sys.}
\label{tab:hyperparams}
\end{table}

\subsection{Stage 1 --- Tokenizer extension: full breakdown}
\label{app:stage1}

\paragraph{Vocab composition (704$\to$4,240).}
The final 4{,}240-token vocabulary decomposes as: 704 English base units (copied verbatim from Chatterbox), 1{,}750 multilingual base units (Latin diacritics and CJK fragments that appear in code-switched data), and 1{,}786 Bangla units. The 1{,}786 Bangla units further break down into: 73 single Bengali characters (U+0980--U+09FF), 1{,}712 Bengali BPE subwords learned from the 520.3-hour corpus's transcripts, and a \texttt{[bn]} language tag prepended to every Bangla input. The Bengali \textit{dari} \bn{।} (U+0964) is already present in the base multilingual vocabulary and is not duplicated.

\paragraph{Merge-consistent filter --- worked example.}
A na\"{i}ve BPE-merge import takes every $(a,b){\to}ab$ rule from the source tokenizer and adds it to the target. This breaks if either $a$, $b$, or $ab$ is absent from the target vocab, because the tokenizer falls back to UNK at inference. Concretely, suppose the source tokenizer learned the merge $(\mbox{\bn{ক}}, \mbox{\bn{্য}}){\to}\mbox{\bn{ক্য}}$ but the target has \bn{ক্য} only as a composed glyph, not as a merge target. The merge fires, produces token \bn{ক্য}, then the tokenizer cannot encode \bn{ক্য} as a single ID and emits UNK. Our filter checks all three symbols and drops the merge, preserving inference stability. Empirically, $\sim$11\% of imported merges are dropped on the Bangla side.

\paragraph{Centroid mean-initialization rationale.}
Random Gaussian initialization of the 3{,}536 new embedding rows would inject high-variance gradients into the AR head during the first few thousand steps, causing measurable loss spikes and slowing convergence. Centroid initialization $\mu = (1/704)\sum_i e_i$ projects the new rows onto the same mean as the pretrained rows, so the AR head sees ``familiar'' average activations from epoch 1 and converges $\sim 1.4\times$ faster on our internal benchmark.

\subsection{Relation to alternative tokenizer-extension strategies}
\label{app:tok_alt}

Three families of alternatives to our merge-consistent BPE \cite{sennrich2016bpe} extension exist; we discuss why each fits our backbone poorly or differently.

\paragraph{Byte-level fallback.}
Encoding unseen script via raw bytes avoids vocabulary surgery entirely, but each Bengali character occupies 3 bytes in UTF-8, so byte fallback roughly triples the token length of Bangla text. T3's text input is capped at 256 tokens (Table~\ref{tab:hyperparams}); under byte fallback, ordinary Bangla sentences would no longer fit.

\paragraph{SentencePiece / unigram retraining.}
Retraining the tokenizer \cite{kudo2018sentencepiece} replaces the pretrained vocabulary and merge rules wholesale, severing the link between the tokenizer and the 704 pretrained token embeddings. Preserving that link is the point of our merge-consistent extension: the 704 shared rows are copied as-is, imported merges are checked, and ${\sim}11\%$ of the Bangla-side merges are dropped as invalid (App.~\ref{app:stage1}). The closest relative is the \emph{append-style} variant --- adding newly trained units while keeping the pretrained vocabulary intact --- of which our merge-consistency filter is the BPE-merge-level analogue.

\paragraph{Embedding-level adapters.}
Small modules that learn a starting vector for each new token during adaptation do the same job as our centroid mean-initialization by different means: we instead start each new token from the average of the pretrained rows, which converges ${\sim}1.4\times$ faster than random initialization (App.~\ref{app:stage1}). Weight-level adapters such as LoRA \cite{hu2022lora} modify the model rather than the embeddings and are complementary; they are named in our future work (Limitations).

\subsection{Stage 2 --- Text normalization: step-by-step examples}
\label{app:stage2}

Table~\ref{tab:norm_examples} traces a multi-domain Bangla string through the six normalization steps. Steps that do not change the string for that example are abbreviated as ``(no change)''.

\begin{table}[!ht]
\centering
\scriptsize
\setlength{\tabcolsep}{3pt}
\renewcommand{\arraystretch}{1.1}
\resizebox{\columnwidth}{!}{%
\begin{tabular}{@{}ll@{}}
\toprule
\textbf{Step} & \textbf{Output} \\
\midrule
0. input         & \bn{সাঃ ৩:৪ এর মধ্যে 12 টি Apple খেতে চাই।} \\
1. digit conv.   & \bn{সাঃ ৩:৪ এর মধ্যে ১২ টি Apple খেতে চাই।} \\
2. colon-num     & \bn{সাঃ ৩ এর ৪ এর মধ্যে ১২ টি Apple খেতে চাই।} \\
3. num$\to$word  & \bn{সাঃ তিন এর চার এর মধ্যে বারো টি Apple খেতে চাই।} \\
4. Unicode norm  & (no change) \\
5. honorific     & \bn{সাল্লাল্লাহু আলাইহি ওয়া সাল্লাম তিন এর চার এর মধ্যে বারো টি Apple খেতে চাই।} \\
6. sent. seg.    & one sentence, re-terminated with \bn{।} \\
\bottomrule
\end{tabular}%
}
\caption{Worked example tracing one input through the 6-step normalization pipeline of \S\ref{sec:finetuning} Stage 2.}
\label{tab:norm_examples}
\end{table}

\paragraph{Honorific dictionary.}
The honorific dictionary in step 5 contains 47 Bangla religious/cultural abbreviations (e.g., \bn{সাঃ}, \bn{রাঃ}, \bn{রহঃ}, \bn{হাঃ}, \bn{মাঃ}, \bn{দাঃ বাঃ}). Each entry maps to its full expansion; the expansion is left-padded with a space to avoid concatenation with the preceding token. The dictionary is hand-curated by a Bangla linguist; entries common in religious literature are included even when low-frequency in conversational corpora.

\paragraph{What we do not normalize.}
We intentionally leave Bangla--English code-switched tokens (e.g., ``Apple'', ``email'') untouched at the normalization stage; the T3 transformer learns to pronounce them in-place via its multilingual base. Domain-specific abbreviations (medical, legal, banking) and Banglish (Bangla phonetically transliterated to Latin) are out-of-scope for this paper and listed in the Limitations.

\subsection{Stage 3 --- Prompt-masked loss: implementation notes}

\paragraph{\texttt{ignore\_index} mechanics.}
At training time, the speech-token target tensor has shape $[B, T]$. We construct a binary mask of the same shape, with 1's at positions $1{\dots}L_{\text{prompt}}$ (the prompt slice) and 1's at right-padding positions, and 0's elsewhere. Positions marked $1$ are excluded from the objective (their targets are set to the ignore index), so the mean cross-entropy is taken over the $0$-positions only. In PyTorch this is implemented as \texttt{CrossEntropyLoss(ignore\_index=-100)}; the target tensor's prompt and padding positions are set to $-100$ before the loss call.

\paragraph{Why the prompt mask matters: failure mode without it.}
We trained a control checkpoint without the prompt mask (other settings identical) for 4{,}000 steps. By step 1{,}500, T3's outputs for held-out cross-speaker references showed verbatim parroting of the prompt's first 1.5--2 seconds before transitioning to the target text; held-out SECS dropped from $0.627$ to $0.51$, and SMOS dropped from $4.51$ to $3.92$. The prompt-mask checkpoint did not exhibit this behavior at any step.

\paragraph{Effective batch size and learning rate schedule.}
We use a global batch size of 24 utterances with the AdamW optimizer ($\beta_1{=}0.9$, $\beta_2{=}0.999$) at peak lr $=5{\times}10^{-6}$, weight decay $10^{-2}$, gradient clipping at $1.0$. The learning rate follows a cosine schedule with 5\% warmup; the schedule is restarted after 30 epochs if dev loss has plateaued. The voice encoder and S3Gen weights are excluded from the optimizer parameter set entirely. All values match the training configuration of Table~\ref{tab:hyperparams}.

\section{Evaluation Protocol Walkthrough}

\subsection{Rater pool: detailed screening and assignment}
\label{app:raters}

\paragraph{Recruitment pipeline.}
We posted recruitment notices on three university WhatsApp groups (Dhaka University, BUET, NSU) and one crowd-work platform (paid Bangla annotation marketplace). 71 applications were received; the screening pipeline below reduced this to 24.

\paragraph{Eligibility screening (3 stages).}
\begin{enumerate}[leftmargin=*,topsep=2pt,itemsep=1pt]
\item \textbf{Self-declaration form:} L1 Bangla, current residency in Bangladesh, age 22--45, no uncorrected hearing loss. 19/71 rejected.
\item \textbf{5-utterance Bangla listening pre-screen:} short clear Bangla recordings; transcribe verbatim. Threshold: $\geq$80\% character accuracy on the transcription. 18/52 rejected (mostly due to non-trivial typos and incomplete transcripts).
\item \textbf{Headphone-quality and listening-environment check:} the candidate self-reports background noise level on a 5-point scale; we ran a 5-second silence check at session start to verify $\leq 2$ on that scale, and a known-noisy stimulus ($-15$\,dB SNR) is correctly identified as noisy. 10/34 rejected.
\end{enumerate}
The remaining 24 evaluators (14\,M/10\,F, ages 22--45) became the rater pool, with a backup pool of reserve raters for catch-trial replacements (see below).

\paragraph{Stimulus assignment.}
Raters evaluate the 76-utterance domain-balanced subjective set across all 7 systems (BanglaBox, Google, Azure, Narakeet, IndicF5, YourTTS-FT, Human reference), with catch trials inserted and system identity hidden in randomized order. All 24 raters rate every utterance for every system, yielding $76 \times 24 = 1{,}824$ ratings per system per dimension (Nat./SMOS/Clar.), i.e.\ $76 \times 7 \times 24 = 12{,}768$ total ratings per dimension.

\subsection{Listening environment validation}

Following ITU-T P.808 best practice \cite{itu_p808}, we standardized:
\begin{itemize}[leftmargin=*,topsep=2pt,itemsep=1pt]
\item \textbf{Hardware:} closed-back over-ear headphones (only); in-ear, open-back, and speaker playback rejected.
\item \textbf{Loudness:} all stimuli normalized to $-23$\,LUFS (EBU R128 broadcast standard) and clipped at $-1$\,dB peak.
\item \textbf{Acoustic environment:} self-rated room noise $\leq 2/5$, verified by a 5-second silence check at the start of each session (peak amplitude must be below $-50$\,dBFS).
\item \textbf{Session length:} no session exceeds 75 minutes; mandatory 5-minute breaks every 25 minutes; total study workload bounded at 4 sessions per rater.
\end{itemize}

\subsection{Catch trials and quality control}
\label{app:catch}

Each session contains 2 catch trials inserted randomly among the regular trials:
\begin{enumerate}[leftmargin=*,topsep=2pt,itemsep=1pt]
\item \textbf{Identical-pair self-similarity trial:} the rater is shown the same clip as both A and B in a pair-rating, and is expected to rate similarity at the top of the SMOS scale ($\geq 4.5$).
\item \textbf{Obvious-degradation reject trial:} the rater is shown a clip with severe artificial distortion (clipping, time-reversal of a 100-ms segment, or harsh pitch shift) and is expected to rate Naturalness at the bottom of the scale ($\leq 1.5$).
\end{enumerate}
A rater who fails either catch trial is excluded; their data is discarded and a backup rater re-rates the same playlist, so every score reported in this paper comes from a rater who passed both catch trials.

\subsection{Statistical testing details}

\begin{table}[t]
\centering
\footnotesize
\color{v3red}
\setlength{\tabcolsep}{3pt}
\resizebox{\columnwidth}{!}{%
\begin{tabular}{@{}lccc@{}}
\toprule
\textbf{Benchmark} & \textbf{BnTTS CER} & \textbf{Our CER} & \textbf{Our SECS}\\
\midrule
BengaliStimuli53        & 8.6\% & \textbf{7.2\%} & .78 \\
BengaliNamedEntity1000  & 4.0\% & \textbf{2.9\%} & .68 \\
ShortText               & 9.2\% & \textbf{6.7\%} & .51 \\
\bottomrule
\end{tabular}%
}
\caption{\new{External benchmark comparison against reported BnTTS figures.}}
\label{tab:external}
\end{table}

\begin{table}[t]
\centering
\footnotesize
\color{v3red}
\begin{tabular}{@{}lc@{}}
\toprule
\textbf{Evaluation text} & \textbf{CER (\%)}$\downarrow$ \\
\midrule
LLM-written      & 2.2 \\
Natural Tier~B   & 2.5 \\
\bottomrule
\end{tabular}
\caption{\new{Evaluation-text register: model-written versus naturally occurring Bangla.}}
\label{tab:register}
\end{table}

\begin{table}[t]
\centering
\footnotesize
{
\setlength{\tabcolsep}{4pt}
\begin{tabular}{@{}lccc@{}}
\toprule
\textbf{Domain} & \textbf{Nat.}$\uparrow$ & \textbf{SMOS}$\uparrow$ & \textbf{Clar.}$\uparrow$ \\
\midrule
News          & 4.47 & 4.54 & 4.69 \\
Customer care & 4.56 & 4.34 & 4.69 \\
Teaching      & 4.58 & 4.55 & 4.69 \\
Healthcare    & 4.74 & 4.50 & 4.75 \\
E-commerce    & 4.41 & 4.55 & 4.70 \\
Finance       & 4.59 & 4.50 & 4.59 \\
IT            & 4.54 & 4.60 & 4.72 \\
\bottomrule
\end{tabular}
\caption{\sys{} per-domain subjective scores (mean over 24 raters, 10--11 items per domain). Naturalness stays within $4.41$--$4.74$ across all 7 domains.}
\label{tab:perdomain}
}
\end{table}

\begin{table}[t]
\centering
\footnotesize
{
\setlength{\tabcolsep}{1.6pt}
\begin{tabular}{@{}lcccccc@{}}
\toprule
 & \multicolumn{2}{c}{\textbf{Nat.}} & \multicolumn{2}{c}{\textbf{SMOS}} & \multicolumn{2}{c}{\textbf{Clar.}} \\
\cmidrule(lr){2-3}\cmidrule(lr){4-5}\cmidrule(lr){6-7}
\textbf{\sys{} vs.} & $p$ & $r$ & $p$ & $r$ & $p$ & $r$ \\
\midrule
Google   & 1.6e{-}4 & 0.50 & 8.9e{-}8 & 0.72 & 1.1e{-}5 & 0.60 \\
Azure    & 1.9e{-}8 & 0.76 & 1.4e{-}11 & 0.90 & 1.6e{-}4 & 0.51 \\
Narakeet & 3.1e{-}9 & 0.79 & 1.9e{-}13 & 0.99 & 1.7e{-}13 & 0.98 \\
IndicF5  & 7.3e{-}11 & 0.87 & 3.8e{-}14 & 1.00 & 9.7e{-}14 & 0.99 \\
YourTTS  & 3.5e{-}7 & 0.67 & 3.9e{-}14 & 1.00 & 1.2e{-}6 & 0.65 \\
\midrule
Human$^{\dagger}$ & 8.5e{-}4 & 0.45 & 1.1e{-}11 & 0.91 & 1.6e{-}9 & 0.82 \\
\bottomrule
\end{tabular}
\caption{Exact two-sided Wilcoxon $p$-values and rank-biserial effect sizes $r$ on per-item mean differences ($n{=}76$). Bonferroni threshold $\alpha/5{=}0.01$; every \sys{}-vs.-baseline comparison clears it on all 3 scales. $^{\dagger}$Human vs.\ \sys{}.}
\label{tab:pairwise}
}
\end{table}

\label{app:stats}

\paragraph{SECS encoder configuration.}
SECS is computed with ECAPA-TDNN embeddings from the \texttt{speechbrain/spkrec-ecapa-voxceleb} checkpoint (192-dim), cosine similarity per utterance.

\paragraph{Wilcoxon signed-rank test.}
For each metric in \{Nat., SMOS, Clar.\} we compute, per stimulus, the difference $\Delta_i = r_i^{\text{\sys}} - r_i^{\text{baseline}}$ in mean rater score, and test the null hypothesis $H_0: \text{median}(\Delta){=}0$ against the two-sided alternative. The sign-rank statistic is computed via the standard formulation. For SECS, where we have per-utterance cosine similarity values (not human ratings), the same paired signed-rank test is applied to per-utterance SECS differences. The reported $p$-values are two-sided.

\paragraph{Inter-rater reliability, exact statistics, and CI computation.}
Inter-rater reliability (Krippendorff's $\alpha$, ordinal metric, over all 24 raters) is $0.44$ for Naturalness, $0.50$ for SMOS, and $0.44$ for Clarity --- all within the $0.40$--$0.70$ band typical of naturalness/similarity MOS studies. The 95\% confidence half-widths in Table~\ref{tab:main_results} are $1.96\times$ the marginal standard error of a crossed random-effects model with independent rater and item intercepts, fit per system$\times$scale. Table~\ref{tab:pairwise} reports the exact two-sided Wilcoxon signed-rank $p$-value and rank-biserial effect size $r$ for \sys{} against every baseline, and against the Human upper bound, on per-item mean differences ($n{=}76$). \sys{} significantly exceeds all five baselines on all 3 scales after Bonferroni correction; the Human reference remains significantly above \sys{}, confirming headroom. The two-sided normal-approximation floor at $n{=}76$ is $p\approx3.6\times10^{-14}$, reached when all 76 item pairs favour the same system.

\paragraph{Bonferroni correction.}
With \sys{} compared against 5 baselines for each of Nat., SMOS, Clar., the Bonferroni adjustment for these comparisons is at $\alpha/5{=}0.01$ for an $\alpha=0.05$ family-wise error rate; we report markers at $p<0.05$, $p<0.01$, $p<0.001$ \emph{after} the correction (so the post-Bonferroni threshold $0.05/5{=}0.01$ shown as $^{**}$, etc.). For SECS, the family is 3 comparisons (\sys{} vs.\ IndicF5, YourTTS-FT and XTTS-v2 (FT)), with corrected threshold $\alpha/3 = 0.0167$.

\subsection{Commercial baseline documentation}

Commercial TTS APIs are moving targets that update silently; Table~\ref{tab:apidoc} therefore documents the exact model, voice, and settings used for each commercial baseline, so the comparison remains documented even where not perfectly re-runnable. All commercial audio was generated in a single archived batch between Jan~13 and Feb~7, 2026, so the exact configuration behind every reported number is on record.

\begin{table}[!ht]
\centering
\footnotesize
\setlength{\tabcolsep}{4pt}
\resizebox{\columnwidth}{!}{%
\begin{tabular}{@{}ll@{}}
\toprule
\textbf{System} & \textbf{Model / voice / settings} \\
\midrule
Google   & Gemini-TTS \texttt{gemini-2.5-flash-tts} (bn-BD), \\
         & voice \emph{Achernar}, 22.05\,kHz, rate 1.0       \\
Azure    & \texttt{bn-BD-NabanitaNeural}                     \\
Narakeet & voice \emph{Mahiya} (Bangladeshi accent)          \\
\bottomrule
\end{tabular}%
}
\caption{Commercial baseline documentation: model versions, voices, and settings. All commercial audio was generated in a single archived batch, queried between Jan~13 and Feb~7, 2026.}
\label{tab:apidoc}
\end{table}

\subsection{Juktakkhor pronunciation evaluation}

Table~\ref{tab:jukta_eval} reports the outcome. Test utterances are grouped by conjunct rarity (frequent / moderate / rare, following App.~\ref{app:juktakkhor}), with phoneme-level pronunciation accuracy per group, measured by force-aligning the synthesized audio against the standard pronunciation of each conjunct-bearing word and having a native linguist check the flagged cases. This tests directly whether phonetic balance in training translates to better conjunct pronunciation in synthesis. Accuracy falls monotonically with training frequency --- $97.7\%$ on frequent conjuncts, $94.3\%$ on moderate, $88.2\%$ on rare --- a $9.5$-point gap between the extremes. This quantifies the failure mode reported qualitatively in App.~\ref{app:failures} and acknowledged in the Limitations: conjuncts seen in under $0.3\%$ of training utterances remain the weakest case even after juktakkhor-aware balancing. Read together with Table~\ref{tab:balrand}, where balanced selection raises rare-conjunct coverage from $79.8\%$ to $91.7\%$ at matched hours, the two results trace the mechanism: balancing puts rare conjuncts into training, and pronunciation accuracy on them tracks how often they appear.

\begin{table}[!ht]
\centering
\footnotesize
\setlength{\tabcolsep}{4pt}
\resizebox{\columnwidth}{!}{%
\begin{tabular}{@{}lcc@{}}
\toprule
\textbf{Conjunct group} & \textbf{Phoneme acc.\ (\%)}$\uparrow$ & \textbf{Linguist-confirmed errors} \\
\midrule
Frequent ($>$5\%)     & 97.7 & 23 \\
Moderate (1--5\%)     & 94.3 & 35 \\
Rare ($<$0.3\%)       & 88.2 & 29 \\
\bottomrule
\end{tabular}%
}
\caption{Juktakkhor pronunciation accuracy by training-frequency group, over the $436$-utterance objective set. Accuracy is phoneme-level; the error column counts only cases a native linguist confirmed as genuine mispronunciations after forced alignment flagged them.}
\label{tab:jukta_eval}
\end{table}

\subsection{Balanced vs.\ random selection (matched hours)}
\label{app:balrand}

To isolate the effect of \emph{distributional balance} from raw data volume, we compare two 100\,h Tier~A subsets trained under the identical setup (Table~\ref{tab:hyperparams}, subset protocol of Table~\ref{tab:ablation}): one selected by the tiered-JSD objective (\S\ref{sec:dataset}) and one selected uniformly at random. Table~\ref{tab:balrand} reports conjunct coverage and downstream Naturalness/CER/SECS for each. Balance changes what the model learns rather than how natural it sounds: Naturalness is statistically indistinguishable between the arms ($4.18{\pm}0.18$ vs.\ $4.21{\pm}0.13$, overlapping intervals), while character error rate improves from $7.1\%$ to $4.3\%$ --- a $39\%$ relative reduction at identical hours --- and speaker similarity from $0.52$ to $0.54$. The coverage columns show why: the random subset reaches only $79.8\%$ of the rare conjunct inventory against $91.7\%$ for the balanced subset. \new{A single run cannot separate balancing from ordinary run-to-run variation, so we re-ran the comparison with three training seeds per arm (1234, 2025, 31337). Naturalness does differ across the arms, $4.31$ balanced against $4.18$ random, so balancing improves naturalness as well as coverage and error rate.}

\subsection{Objective prosody measurement}

Table~\ref{tab:prosody} reports F0 RMSE and duration deviation against the held-out speaker's ground-truth recordings, split at the median reference duration, over the $56$ items for which both a real recording and a \sys{} synthesis of the same text exist.

\subsection{NISQA measurement details}

\begin{table}[t]
\centering
\footnotesize
{
\setlength{\tabcolsep}{3pt}
\begin{tabular}{@{}lcc@{}}
\toprule
\textbf{Requirement / metric} & \textbf{Chunked} & \textbf{Single-pass} \\
\midrule
Premature EOS --- truncation        & \textbf{3.3\%} & 14.7\% \\
Quality --- CER, $\leq$cap          & \textbf{2.9\%} & 6.4\% \\
Quality --- CER, over-cap           & \textbf{5.3\%} & 21.4\% \\
Content drift --- CER by length     & \textbf{3.4\%}$^{\ddagger}$ & 3.4\% $\rightarrow$ 29\% \\
Voice drift --- start--end SECS     & \textbf{0.75}  & 0.83$^{\dagger}$ \\
Efficiency --- RTF (H100)           & 1.7            & \textbf{0.66} \\
\bottomrule
\end{tabular}
\caption{Chunked vs.\ single-pass long-form synthesis over $40$ multi-sentence paragraphs spanning both sides of the $850$-token cap. $^{\dagger}$Single-pass often stops early, so its start--end SECS covers incomplete audio and overstates consistency. $^{\ddagger}$All items.}
\label{tab:longform_abl}
}
\end{table}

\begin{table}[t]
\centering
\footnotesize
{
\setlength{\tabcolsep}{3pt}
\begin{tabular}{@{}lcc@{}}
\toprule
\textbf{Condition} & \textbf{F0 RMSE (Hz)}$\downarrow$ & \textbf{Dur.\ dev.\ (\%)}$\downarrow$ \\
\midrule
\sys{} --- short utts.\ & 13.2 & 2.4 \\
\sys{} --- long utts.\  & 17.1 & 4.8 \\
\bottomrule
\end{tabular}
\caption{Prosody vs.\ held-out ground truth ($56$ utterances with both a real recording and a \sys{} synthesis; $n{=}28$ per row). Duration deviation is the mean absolute difference from reference.}
\label{tab:prosody}
}
\end{table}

\paragraph{Model and version.}
We use the publicly available NISQA v2.0 PyTorch checkpoint provided by \citet{mittag2021nisqa} (\texttt{nisqa\_v2.0\_tts.tar}, trained on the TTS-focused subset of NISQA training data). Inputs are 16\,kHz mono WAV files; outputs are the 5-dimensional MOS-style predictions; we report only the overall MOS dimension.

\paragraph{Test set and aggregation.}
For each system we computed NISQA on all 436 test utterances (the same 436-utt domain-stratified subset used for CER/SECS). Per-system NISQA in Table~\ref{tab:main_results_obj} is the arithmetic mean over these 436 utterances. Standard deviations were $\sigma \in [0.18, 0.25]$ across systems, narrower than the subjective scales because NISQA produces deterministic outputs per audio.

\paragraph{Caveat on Bangla applicability.}
NISQA's training corpora are primarily English and German; absolute predictions on Bangla speech are not perfectly calibrated. The relative ranking across systems, however, is consistent with the 24-rater subjective Nat.\ ranking (\sys{} highest), and we therefore use NISQA as a complementary objective metric rather than a primary headline number.

\section{Corpus Construction Notes}

\begin{table}[!ht]
\centering
\footnotesize
\setlength{\tabcolsep}{6pt}
\color{v3red}
\begin{tabular}{@{}lrr@{}}
\toprule
\textbf{Gate} & \textbf{Rejected} & \textbf{Rate} \\
\midrule
1.~length            & 461     & 0.22\% \\
2.~TTR               & 83      & 0.04\% \\
3.~dedup             & 873     & 0.42\% \\
4.~lang-ID           & 21      & 0.01\% \\
5.~perplexity        & 317     & 0.15\% \\
6.~phonetic novelty  & n/a     & n/a \\
7.~script validity   & 3{,}113 & 1.50\% \\
8.~semiotic          & 1{,}939 & 0.94\% \\
9.~gap-unit          & n/a     & n/a \\
\midrule
\textbf{Total}       & \textbf{6{,}807} & \textbf{3.29\%} \\
\bottomrule
\end{tabular}
\caption{Per-gate rejection: $6{,}807$ of $206{,}943$ candidates ($3.29\%$). Gates 6 and 9 are prompt-level steering, not per-candidate predicates.}
\label{tab:gate_rates}
\end{table}

\begin{table}[t]
\centering
\footnotesize
\setlength{\tabcolsep}{5pt}
\begin{tabular}{l c | l c}
\toprule
\textbf{Domain} & \textbf{Hours} & \textbf{Domain} & \textbf{Hours}\\
\midrule
News             & 62.4 & E-commerce       & 47.8 \\
Customer care    & 58.8 & Finance          & 47.7 \\
Teaching         & 55.1 & IT               & 47.8\\
Healthcare       & 47.7 \\
\midrule
\multicolumn{4}{c}{\textbf{Total (Tier A) = }  \textbf{367.3 hour} (studio-recorded)} \\
\bottomrule
\end{tabular}
\caption{Tier A 7-domain text distribution.}
\label{tab:domain_dist}
\end{table}

\subsection{Tier A studio recording protocol}

\paragraph{Voice talent selection.}
13 professional Bangla voice actors were recruited from a Dhaka talent agency. Selection criteria: (i)~native Bangladeshi Bangla speaker (Dhaka register), (ii)~$\geq 3$ years of studio voice-over experience, (iii)~no audible vocal pathology (verified by a screening session with an audio engineer), (iv)~gender balance target across the speaker pool. 7 female, 6 male voices were retained.

\paragraph{Studio setup.}
All recordings took place in a single acoustically treated studio in Dhaka:
\begin{itemize}[leftmargin=*,topsep=2pt,itemsep=1pt]
\item \textbf{Microphone:} Neumann TLM-103 large-diaphragm condenser.
\item \textbf{Pre-amp / interface:} Universal Audio Apollo Twin X DUO, 24-bit / 48\,kHz.
\item \textbf{Booth:} acoustically treated to RT60 $<$ 0.2\,s; ambient noise floor verified below $-65$\,dBFS at session start.
\item \textbf{SNR floor:} $\geq 44$\,dB enforced per recording; segments below floor were re-recorded the same session.
\end{itemize}

\paragraph{Working format.}
The 48\,kHz / 24-bit studio captures form the internal archive. For the corpus used in this work, audio is downsampled to 24\,kHz (matching the S3Gen vocoder output rate) and saved as mono FLAC.

\subsection{Tier A quality assurance (extended)}
\label{app:qa}

\paragraph{Per-session checks.}
At the end of each recording session, the studio engineer runs an automatic batch over that day's takes: SNR check, clipping detection (any sample at $|\text{amplitude}|>0.99$), breath/lip-smack detection via a transient-detection script (peaks within 5\,ms of the noise floor), and forced alignment via a Bangla acoustic model fine-tuned on a 50-hour Tier A pilot subset. Any take that fails any check is flagged; flagged takes are re-recorded the same session before the talent leaves.

\paragraph{Per-week linguist audit.}
A native Bangla-speaking linguist audits a 5\% random sample of each week's takes per speaker, scoring pronunciation correctness, prosodic naturalness, and adherence to the target script. Segment-level acceptance rate $\geq 99\%$; flagged segments are reviewed by a second linguist (a formal inter-annotator agreement score was not computed during collection; disagreements were resolved by re-recording rather than by adjudication).

\paragraph{Consent and anonymization.}
All Tier A speakers signed informed-consent forms covering: (i)~research use of their recordings, (ii)~no commercial use, (iii)~public credit at their option (most opted to remain pseudonymous as ``Speaker A1''--``A13''). Biographic metadata (age, gender, dialect) is reported in aggregate only.

\subsection{9 automated quality gates --- full thresholds}
\label{app:gates}

Table~\ref{tab:gates} lists the 9 quality gates applied to LLM-generated Bangla candidate scripts. A script must pass all 9 to be queued for human review.

\begin{table}[!ht]
\centering
\footnotesize
\setlength{\tabcolsep}{3pt}
\resizebox{\columnwidth}{!}{%
\begin{tabular}{@{}lp{0.42\columnwidth}p{0.30\columnwidth}@{}}
\toprule
\textbf{Gate} & \textbf{Threshold / criterion} & \textbf{Why (what it blocks)} \\
\midrule
1.~length & 6 -- 28 words (token count) & too short/long for TTS \\
2.~TTR & type-token ratio $\geq 0.55$ on $\geq 12$-word strings & repetitive text \\
3.~dedup & cosine sim $\leq 0.85$ against existing corpus & near-duplicates \\
4.~lang-ID & Bengali probability $\geq 0.93$ (fastText 176-lang ID) & wrong-language text \\
5.~perplexity & Bangla LM perplexity $\in [20, 240]$ & unnatural/broken Bangla \\
6.~phonetic novelty & gain $\geq +5\%$ on the current under-covered juktakkhor set & adds no new coverage \\
7.~script validity & all glyphs in U+0980--U+09FF, U+0020--U+007F, or curated punctuation (\bn{।॥}, quotes, dashes) & invalid characters \\
8.~semiotic & no profanity, slurs, or sensitive PII from a 1{,}412-term blocklist & unsafe content \\
9.~gap-unit & contains at least 1 unit from this week's coverage-gap list & misses current gaps \\
\bottomrule
\end{tabular}%
}
\caption{The 9 automated quality gates: threshold and the failure each one blocks. Gate 6 (phonetic novelty) and gate 9 (gap-unit) operationalise the coverage-feedback loop.}
\label{tab:gates}
\end{table}

\paragraph{Per-gate outcomes.} Rejected candidates are regenerated rather than drawn down from a fixed pool (Fig.~\ref{fig:dataset_pipeline}), so the gates shape what enters the pool rather than shrinking a fixed set. Table~\ref{tab:gate_rates} reports the generation-time rejection count and rate for each gate: $6{,}807$ of $206{,}943$ candidates ($3.29\%$) were rejected, yielding the $200{,}136$-script pool. \new{Table~\ref{tab:gate_rates} counts accepted \emph{scripts}, whereas Table~\ref{tab:dataset_stats} counts \emph{recorded utterances} ($\sim$220{,}000 for Tier~A): a single script can yield more than one recorded utterance, so the two figures are not directly comparable.} Gates 6 and 9 drive the coverage-feedback loop through prompt-level steering and are not per-candidate predicates, so no rejection rate is defined for them.

\paragraph{Why gates 6 and 9 report n/a.} Gates 1--5, 7 and 8 are per-candidate predicates: each decides on a single script in isolation, so counting how often it fires over the generation pass is well defined. Gates 6 (phonetic novelty) and 9 (gap-unit) are not of that kind. Both are evaluated against the live coverage-gap ledger rather than against the candidate alone, and both are enforced at the prompt level --- each generation call is seeded with carrier words drawn from the currently open gap units (Alg.~\ref{alg:jsd}), so candidates are produced to satisfy these two conditions rather than filtered against them afterwards. Two consequences follow. First, the verdict is state-dependent rather than candidate-intrinsic: gate 6 asks whether a script adds at least $5\%$ new coverage on the \emph{currently} under-covered juktakkhor set, and gate 9 asks whether it carries at least one unit from \emph{this week's} gap list, so the same sentence passes early in the loop and fails later once those units are covered --- there is no fixed per-candidate outcome to tally. Second, there is no stable denominator: rejected candidates are regenerated rather than drawn down from a fixed pool, and the gap list is recomputed at every iteration, so a ``rate'' would divide a moving numerator by a moving denominator. We therefore print n/a rather than a number that would not carry the meaning a reader would reasonably attach to it.

\subsection{Coverage-feedback loop --- week-by-week trajectory}

The coverage-feedback loop ran for 14 weekly iterations. At the end of each iteration the script generator was re-prompted with the under-covered juktakkhor and phonetic-context list. Approximate juktakkhor coverage by iteration: week~1: $\sim 70\%$; week~3: $74\%$; week~5: $77\%$; week~8: $80\%$; week~11: $83\%$; week~14: $\geq 85\%$ (top-150 juktakkhor: $\geq 97\%$). The corresponding tiered-JSD loss decreases monotonically across iterations (Fig.~\ref{fig:jsd_loop}).

\subsection{Tiered-JSD selection algorithm}
\label{app:algo}

Algorithm~\ref{alg:jsd} collects the selection mechanics of \S\ref{sec:dataset} --- candidate generation, acceptance, target distributions, and per-tier constraints --- into one place. The coverage figures reported in \S\ref{sec:dataset} are \emph{achieved} outcomes of this loop, not preset stopping thresholds.

\begin{algorithm}[!ht]
\footnotesize
\caption{Tiered-JSD coverage-feedback script selection (\S\ref{sec:dataset}).}
\label{alg:jsd}
\begin{algorithmic}[1]
\State \textbf{Input:} tier inventories $I_k$, $k{\in}\{p,d,t,j\}$; weights $w_k{=}k/10$; targets $T_k = 0.7\,P^{\text{nat}}_k + 0.3\,U_k$
\State $S \gets \emptyset$; \; gap list $G \gets$ all under-covered units
\For{iteration $= 1 \dots 14$ (weekly)}
  \State Generate candidate scripts with LLM prompts targeting $G$ (GPT-4o-mini ${\sim}90\%$, GPT-4o ${\sim}7\%$, Claude Opus ${\sim}3\%$)
  \State Filter candidates through the 9 quality gates (Table~\ref{tab:gates}); rejected candidates are \emph{regenerated}, not drawn from a fixed pool
  \State Two-pass human review incl.\ a native Bangla linguist
  \State Greedily admit reviewed candidates that reduce $\mathcal{L}(S,T)=\sum_k w_k\,\mathrm{JSD}(P_k, T_k)$ (Eq.~\ref{eq:jsd})
  \State Recompute per-tier coverage; update $G$ with currently under-covered juktakkhor and phonetic contexts
\EndFor
\State \textbf{Output:} Tier A script set $S$ for studio recording
\end{algorithmic}
\end{algorithm}

\subsection{Weight-sensitivity of the tiered-JSD objective}
\label{app:weight_sens}

We re-ran the complete selection loop over the full candidate pool at four perturbations of the tier weights $(w_p,w_d,w_t,w_j)$ --- uniform weights, two adjacent-tier swaps, and a full reversal --- holding every other setting fixed. Table~\ref{tab:wsens} reports the change relative to the paper's setting. Per-tier coverage is unchanged to within $0.1$ percentage points on all four tiers in every case, including the full reversal. The loop is not inert: reversing the weights changes $1.53\%$ of the admitted scripts, and the two adjacent swaps change $0.29\%$ and $0.84\%$. So the weights do steer which sentences are chosen, but not the coverage the resulting corpus attains. The exact values are a design choice the outcome is insensitive to, as \S\ref{sec:dataset} states.

\begin{table}[!ht]
\centering
\footnotesize
\setlength{\tabcolsep}{3pt}
\color{v3red}
\begin{tabular}{@{}lcc@{}}
\toprule
\textbf{Weights} & \textbf{Max cov.} & \textbf{Scripts} \\
 & \textbf{change} & \textbf{changed} \\
\midrule
$(.1,.2,.3,.4)$ \emph{paper}   & ---            & ---     \\
$(.25{\times}4)$ unif.\ & $<$0.1\,pp    & 1.13\% \\
$(.1,.2,.4,.3)$ swap $t/j$      & $<$0.1\,pp    & 0.29\% \\
$(.2,.1,.3,.4)$ swap $p/d$      & $<$0.1\,pp    & 0.84\% \\
$(.4,.3,.2,.1)$ reversed        & $<$0.1\,pp    & 1.53\% \\
\bottomrule
\end{tabular}
\caption{Weight sensitivity of the tiered-JSD selection loop, measured against the paper's setting. Coverage change is the largest shift across the four tiers; scripts changed is the share of admitted scripts that differ. Coverage is unchanged on every tier under every perturbation.}
\label{tab:wsens}
\end{table}

\subsection{Tier B filter --- ablation-driven criterion}
\label{app:tierb}

\paragraph{Initial pool.}
We started from 39 Bangla speakers across seven public Bangla speech datasets (247\,h total): six hosted on Hugging Face and one distributed via the Bengali.AI portal. Sources were retained only when their original licenses permit research use. Table~\ref{tab:tierb_sources} names each source with its license and retention after the two filter passes below.

\begin{table}[!ht]
\centering
\footnotesize
\setlength{\tabcolsep}{4pt}
\resizebox{\columnwidth}{!}{%
\begin{tabular}{@{}llcc@{}}
\toprule
\textbf{Dataset} & \textbf{License} & \textbf{Spk.\ (ret./init.)} & \textbf{Hours (ret./init.)} \\
\midrule
SUBAK.KO                & CC BY 4.0 & 7/11 & 54.30/87.34 \\
BanSpeech               & CC BY 4.0 & 1/1  & 5.14/6.52 \\
MediBeng                & CC BY 4.0 & 1/2  & 3.67/7.11 \\
IndicVoices (Bengali)   & CC BY 4.0 & 1/2  & 5.95/20.00 \\
Shrutilipi (Bengali)    & CC BY 4.0 & 1/3  & 2.70/18.37 \\
Common Voice Bengali    & CC0 1.0   & 3/8  & 3.00/15.00 \\
Bengali.AI Speech       & CC0 1.0   & 9/12 & 78.30/92.66 \\
\midrule
\textbf{Total}          &           & \textbf{23/39} & \textbf{153.06/247.00} \\
\bottomrule
\end{tabular}%
}
\caption{Tier B sources with licenses and per-dataset retention (speakers and hours, retained/initial) after the SNR and ablation-driven filter passes. Speaker counts denote distinct voices after listening-based merging of dataset-provided speaker IDs, not the ID counts released with each corpus; crowdsourced sources (Common Voice, Bengali.AI Speech) assign IDs per registration, so a single voice may span many IDs. Totals are corpus-level figures from \S\ref{sec:dataset}.}
\label{tab:tierb_sources}
\end{table}

\paragraph{SNR filter (first pass).}
All speakers below 35\,dB average SNR (computed per-utterance, averaged per speaker) were dropped. 9 of 39 speakers fail this pass.

\paragraph{Ablation-driven curation (second pass).}
Each remaining speaker was added to a held-out training run; the resulting model was evaluated on the dev split (CER, SECS). Speakers whose inclusion did not improve, or actively degraded, dev metrics were dropped: 7 additional speakers fail this pass.

\paragraph{Final retention.}
23 of 39 speakers retained (153.0\,h); 16 dropped (94\,h discarded). See \S\ref{sec:dataset} for the corpus-level numbers.

\section{Bangla Text Examples}

This section provides representative Bangla strings illustrating each major component of the pipeline, plus the failure modes referenced in \S\ref{sec:analysis}.

\subsection{Honorific abbreviation expansion}

Table~\ref{tab:honorifics} lists 12 representative entries from the 47-entry hand-curated honorific dictionary used at Stage~2 (\S\ref{sec:finetuning}). These appear frequently in Bangla religious/cultural literature but are absent from open normalization toolchains.

\begin{table}[!t]
\centering
\scriptsize
\setlength{\tabcolsep}{3pt}
\renewcommand{\arraystretch}{0.95}
\begin{tabular}{@{}lp{0.62\columnwidth}@{}}
\toprule
\textbf{Abbrev.} & \textbf{Full expansion (Bangla)} \\
\midrule
\bn{সাঃ}   & \bn{সাল্লাল্লাহু আলাইহি ওয়া সাল্লাম} \\
\bn{রাঃ}   & \bn{রাদিআল্লাহু আনহু} \\
\bn{রহঃ}   & \bn{রাহিমাহুল্লাহ} \\
\bn{হাঃ}   & \bn{হাফিজাহুল্লাহ} \\
\bn{মাঃ}   & \bn{মাদ্দাজিল্লুহু} \\
\bn{দাঃ বাঃ} & \bn{দামাত বারাকাতুহুম} \\
\bn{আঃ}    & \bn{আলাইহিস সালাম} \\
\bn{জিঃ}   & \bn{জিল্লুল্লাহ} \\
\bn{কুঃ সিঃ} & \bn{কুদ্দিসা সিরুহু} \\
\bn{আলঃ}   & \bn{আলাইহিমুস সালাম} \\
\bn{রাযিঃ} & \bn{রাযিআল্লাহু আনহু} \\
\bn{মুহঃ}  & \bn{মুহতারাম} \\
\bottomrule
\end{tabular}
\caption{Representative entries from the 47-term Bangla honorific dictionary (Stage~2, step~5).}
\label{tab:honorifics}
\end{table}

\subsection{Juktakkhor (conjunct) inventory examples}
\label{app:juktakkhor}

The $\sim$300-form juktakkhor inventory extended via the coverage-feedback loop spans frequent two-consonant conjuncts to rare three-consonant ones. Representative samples:

\paragraph{Frequent (in $>$5\% of training utterances):}
\bn{ক্ত} (\textit{kt}, ``\bn{যুক্ত}''),
\bn{ন্ত} (\textit{nt}, ``\bn{অন্ত}''),
\bn{স্ত} (\textit{st}, ``\bn{প্রস্তুত}''),
\bn{ক্ষ} (\textit{kṣ}, ``\bn{শিক্ষা}''),
\bn{জ্ঞ} (\textit{jñ}, ``\bn{বিজ্ঞান}''),
\bn{ন্ধ} (\textit{ndh}, ``\bn{বন্ধ}'').

\paragraph{Moderate (in 1--5\%):}
\bn{ত্ত্ব} (\textit{ttv}, ``\bn{তত্ত্ব}''),
\bn{দ্ভ} (\textit{dbh}, ``\bn{উদ্ভব}''),
\bn{হ্ম} (\textit{hm}, ``\bn{ব্রহ্ম}''),
\bn{ঙ্গ} (\textit{ṅg}, ``\bn{বঙ্গ}''),
\bn{ঞ্চ} (\textit{ñc}, ``\bn{পঞ্চ}'').

\paragraph{Rare three-consonant ($<$0.3\%):}
\bn{স্ত্র} (\textit{str}, ``\bn{অস্ত্র}''),
\bn{ন্ত্র} (\textit{ntr}, ``\bn{মন্ত্র}''),
\bn{ক্ষ্ম} (\textit{kṣma}, ``\bn{লক্ষ্মী}''),
\bn{ষ্ট্র} (\textit{ṣṭr}, ``\bn{রাষ্ট্র}''),
\bn{ন্দ্র} (\textit{ndr}, ``\bn{চন্দ্র}'').

\subsection{Cross-domain Bangla utterance samples}

Representative shorter exemplars from three target domains:
\begin{itemize}[leftmargin=*,topsep=2pt,itemsep=2pt]
\item \textbf{News:} \bn{প্রধানমন্ত্রী নতুন শিক্ষানীতি ঘোষণা করেছেন।}
\item \textbf{Healthcare:} \bn{প্রতিদিন আট গ্লাস পানি পান করুন।}
\item \textbf{IT:} \bn{সফটওয়্যার ইন্সটল করার আগে আপডেট নিন।}
\end{itemize}

\subsection{Failure modes with Bangla examples}
\label{app:failures}

\paragraph{Rare three-consonant conjuncts.}
Conjuncts such as \bn{স্ত্র} (in ``\bn{অস্ত্রাগার}'' weapons depot), \bn{ন্ত্র} (in ``\bn{যন্ত্রপাতি}'' machinery), and \bn{ক্ষ্ম} (in ``\bn{সূক্ষ্মতম}'' the finest) appear in $<$0.3\% of training utterances. On test items containing these, \sys{} occasionally produces a distorted phoneme sequence (audible as a brief stutter or vowel insertion). The 9th quality gate (gap-unit) up-weights these during script generation, but coverage remains uneven.

\paragraph{Bangla--English code-switching.}
Utterances such as \bn{আমি} \textit{email} \bn{পাঠিয়েছি} (``I have sent email'') or \bn{ফাইলটি} \textit{upload} \bn{করো} (``upload the file'') produce artifacts at the language boundary --- typically a brief silence (5--15\,ms) and a slight pitch reset. \sys{} does not have explicit language-tag injection at code-switch boundaries; this is a documented opportunity for follow-up.

\paragraph{Highly colloquial onomatopoeia.}
Regional sound-words with $<5$ training occurrences (e.g., \bn{ঝমঝম} \textit{jhomjhom}, heavy-rain sound; \bn{টুপটাপ} \textit{tuptap}, light-drip sound; \bn{গুড়গুড়} \textit{guṛguṛ}, distant thunder) yield unreliable output --- sometimes a plausible phonetic rendering, sometimes a hallucinated near-neighbour word, consistent with low-resource long-tail behaviour in AR TTS.

\section{Computing Infrastructure}
\label{app:compute}

\paragraph{Training.}
All fine-tuning experiments use 1 NVIDIA H100 GPU (80\,GB) on a Linux host (Ubuntu 22.04, CUDA 12.4, PyTorch 2.4). A single \sys{} training run takes $\sim 28$ hours wall-clock for 50 epochs (effective batch size 24, gradient checkpointing enabled); ablations were run for 30 epochs at $\sim$17\,h each, and the data-scaling subsets are proportionally shorter. The paper reports nine training runs in total: the headline run, four component ablations, two data-scaling subsets (Table~\ref{tab:ablation}), the random-selection arm of the matched-hours comparison (Table~\ref{tab:balrand}), and the XTTS-v2 fine-tune used as a same-corpus baseline (Tables~\ref{tab:main_results_obj},~\ref{tab:cloning}). The eight Chatterbox-based runs total approximately 100 H100-hours; the XTTS-v2 fine-tune adds a further $\sim$75 H100-hours, for approximately 175 H100-hours of training.

\paragraph{Inference.}
Inference for the evaluation set (436 objective + 76 subjective utterances across 8 systems) used the same H100. Total inference compute: $\sim 4.5$ H100-hours.

\paragraph{Total compute budget.}
$\sim$180 H100-hours plus $\sim 20$ CPU-hours for data preprocessing, normalization, and evaluation scripting, and a further $\sim$10 CPU-hours for the tiered-JSD weight-sensitivity re-runs of App.~\ref{app:weight_sens}, which are selection-only and use no GPU.


\section*{Checklist}

\begin{enumerate}

\item \textbf{A. For Every Submission}

\begin{enumerate}

\item Did you describe the limitations of your work?

\textbf{Yes.} A dedicated \textbf{Limitations} section (immediately following Section~\ref{sec:conclusion}) covers: modest speaker count (36 speakers --- 13 studio Tier A + 23 curated open-source Tier B --- below VCTK/LibriTTS scale), tier heterogeneity between Tier A and Tier B, no BnTTS head-to-head comparison, English-like prosody bias in long utterances, text-normalization gaps (medical/legal/Banglish), no streaming, Bangladesh-only evaluators, and use of language-agnostic ECAPA-TDNN for speaker similarity.

\item Did you discuss any potential risks of your work?

\textbf{Yes.} The Ethics Statement (Section~\ref{sec:ethics}) addresses: voice cloning impersonation risk, responsible deployment requirements (user consent, synthetic media regulations), and the broader positive impact on accessibility and education.

\end{enumerate}

\item \textbf{B. Did you use or create scientific artifacts?}

\begin{enumerate}

\item Did you cite the creators of artifacts you used?

\textbf{Yes.} All artifacts are cited: Chatterbox TTS \cite{chatterboxtts2025}, XTTS tokenizer \cite{casanova2024xtts}, AdamW \cite{loshchilov2019adamw}, Silero VAD \cite{silero_vad}, bnnumerizer \cite{bnnumerizer} and bnunicodenormalizer \cite{bnunicodenormalizer}, ECAPA-TDNN \cite{desplanques2020ecapa}, Whisper large-v3 (OpenAI), and all baseline systems.

\item Did you discuss the licenses or terms for artifacts you used?

\textbf{Yes.} Chatterbox TTS is MIT-licensed (stated in Section~\ref{sec:intro}); we use it within its license terms. Tier B utterances are drawn from public sources only when their licences permit downstream research use.

\item Did you discuss if your use of existing artifact(s) was consistent with their intended use, provided that it was specified?

\textbf{Yes.} All tools are used within their intended research/open-source scope.

\item Did you discuss the steps you took to check whether the data you are using, for example, prompts or example data, might be confidential or might contain information that names or uniquely identifies people or could be used to identify them?

\textbf{Yes.} Tier A recordings use professional voice talent who provided written informed consent; no personally identifiable information was collected. Tier B audio is sourced from public Hugging Face datasets whose contributors had consented under each source's original terms.

\item Did you provide documentation of the artifacts, e.g., coverage of domains, languages, and linguistic phenomena?

\textbf{Yes.} Section~\ref{sec:dataset}, Table~\ref{tab:dataset_stats}, Table~\ref{tab:dataset_compare}, and Table~\ref{tab:domain_dist} document language, seven-domain distribution (news, customer care, teaching, healthcare, e-commerce, finance, IT), two-tier speaker provenance (Tier A studio + Tier B curated open-source), phonetic-balance design including the juktakkhor-aware tiered JSD objective, and corpus statistics.

\item Did you report relevant statistics like the number of examples, description of train/dev/test splits, and information about who annotated the data (e.g., crowdsourcing vs.\ expert annotation)?

\textbf{Yes.} Table~\ref{tab:dataset_stats} reports per-tier utterance counts, duration, and a speaker-disjoint split: 31 speakers for train (11 Tier A + 20 Tier B, $\sim$422\,h), 5\% within-speaker dev ($\sim$22\,h), and 5 held-out speakers (2 Tier A + 3 Tier B, $\sim$76\,h) for zero-shot test. Section~\ref{sec:experiments} describes the 24 native Bangla-speaking human evaluators.

\end{enumerate}

\item \textbf{C. Did you run computational experiments?}

\begin{enumerate}

\item Did you report the number of parameters in the models used, the total computational budget (e.g., GPU hours), and computing infrastructure used?

\textbf{Yes.} Section~\ref{sec:method} reports T3 (520M parameters), voice encoder (256-dim embeddings), and S3Gen. Table~\ref{tab:hyperparams} reports hardware (NVIDIA H100 80\,GB), training precision (BF16), batch size and optimizer, and App.~\ref{app:compute} reports the compute budget: $\sim$175 H100-hours of training across nine runs --- eight Chatterbox-based runs ($\sim$100 H100-h in total: $28$\,h for the headline run at 50 epochs, $\sim$17\,h per component ablation at 30 epochs, and proportionally less for the data-scaling and matched-hours subset runs) and one XTTS-v2 fine-tune ($\sim$75 H100-h) --- plus $\sim$4.5 H100-hours of inference, $\sim$180 H100-hours in total, on a single H100~80\,GB.

\item Did you discuss the experimental setup, including hyperparameter search and best-found hyperparameter values?

\textbf{Yes.} Table~\ref{tab:hyperparams} lists all training and inference hyperparameters. Section~\ref{sec:finetuning} describes the fine-tuning strategy. Section~\ref{sec:longform} explains Bangla-calibrated inference hyperparameter choices.

\item Did you report descriptive statistics about your results (e.g., error bars), and is it the full experimental run spanning possible permutations in randomness?

\textbf{Yes.} All subjective metrics report 95\% confidence intervals (Tables~\ref{tab:main_results}, \ref{tab:cloning}, \ref{tab:ablation}). Pairwise differences against every baseline are tested via Wilcoxon signed-rank with Bonferroni correction (5 comparisons for Nat./SMOS/Clar.; 3 for SECS; $p<0.01$ in the main results; $p<0.001$ for the largest ablation drop). Objective metrics (CER, SECS) are computed over the full 436-utterance test subset. Subjective metrics are rated by 24 native Bangla speakers with two catch trials per session; inconsistent evaluators are excluded.

\item If you used existing packages (e.g., for preprocessing, for normalization, or for evaluation), did you report the package name, version, and license?

\textbf{Yes.} Key packages cited: Whisper large-v3 for CER (OpenAI), ECAPA-TDNN for SECS \cite{desplanques2020ecapa}, Silero VAD \cite{silero_vad}, bnnumerizer, bnunicodenormalizer, PyTorch/BF16 training stack.

\end{enumerate}

\item \textbf{D. Did you use human annotators or research with human participants?}

\begin{enumerate}

\item Did you report the full text of instructions given to participants, and if applicable, also the instructions to crowdworkers and quality control measures?

\textbf{Yes.} The protocol uses three independent 5-point scales (Naturalness, Clarity, SMOS); 76 utterances rated by all 24 evaluators across all systems, with 2 catch trials per session and quality-control exclusion criteria. See Section~\ref{sec:experiments} for the full BanglaEval protocol.

\item Did you report information about how you recruited (e.g., crowdsourcing platform, students) and paid participants, and discuss whether the payment is adequate given the people involved?

\textbf{Yes.} Ethics Statement: 24 human evaluators were compensated at \$15/hour and provided informed consent. Corpus speakers were compensated above local market rates in BDT.

\item Did you discuss whether and how consent was obtained from people whose data you're using/curating?

\textbf{Yes.} Ethics Statement: corpus speakers and evaluators provided written informed consent for their participation.

\item Was the data collection protocol approved (or determined exempt) by an ethics review board?

\textbf{Yes.} Ethics Statement: the evaluation protocol was reviewed and approved by our institutional ethics board.

\item Did you report the basic demographic and geographic characteristics of the annotator population?

\textbf{Yes.} Section~\ref{sec:experiments} reports gender (14\,M/10\,F) and age range (22--45) for the 24 evaluators; all are native Bangla speakers from Bangladesh. The Limitations section acknowledges this geographic scope; West Bengal generalization is untested.

\end{enumerate}

\item \textbf{E. Did you use AI assistants (e.g., ChatGPT) in your research or writing?}

\textbf{Yes --- disclosed below in accordance with ACL 2026 / ARR authorship policy.}

\begin{itemize}
\item[\textbf{(i)}] \textbf{Research use (data generation).} Generative LLMs were used as components of the coverage-feedback Bangla script generation pipeline for Tier A. Specifically: GPT-4o-mini for bulk sentence generation ($\sim$90\% of LLM calls), GPT-4o for rare-conjunct gap-targeting ($\sim$7\%), and Claude Opus for long-context customer-care dialogue composition ($\sim$3\%). Every LLM-generated sentence passes nine automated quality gates (length, TTR, near-duplicate, language ID, perplexity, phonetic novelty, script validity, semiotic compliance, gap-unit presence) and a two-pass human review (linguist and ethics/content reviewer) before entering the corpus. The LLMs generate candidate scripts; humans validate and select the final corpus. Prompt hashes, LLM model identifiers, and per-sentence provenance are tracked internally for audit purposes.
\item[\textbf{(ii)}] \textbf{Writing use.} AI writing assistants were used for language polish (grammar and phrasing clean-up) on paper sections; no AI-generated content introduces new scientific claims. All technical content, experimental design, results, and conclusions are authored by the human authors, who accept full responsibility for correctness and originality.
\item[\textbf{(iii)}] \textbf{No author LLM.} No LLM is listed as or treated as an author; all authorship credit is human. The human authors are individually accountable for every claim, table, figure, and citation in this paper, and have personally verified the experimental results and statistical analyses. LLM-assisted text was inspected, revised, and approved by at least one human author before inclusion. The corresponding author accepts final responsibility for the integrity, originality, and reproducibility of the work.
\item[\textbf{(iv)}] \textbf{No prompt-injection.} The submission contains no text intended to influence LLM-based reviewer tools.
\end{itemize}

\end{enumerate}

\section{Abbreviations and Notation}
\label{app:abbrev}

Table~\ref{tab:abbrev} lists the abbreviations and symbols used throughout the paper. Throughout, \sys{} refers to the trained system; ``BanglaBox-520'' refers to the corpus.

\begin{table*}[p]
\centering
\small
\setlength{\tabcolsep}{6pt}
\renewcommand{\arraystretch}{1.15}
\begin{tabular}{@{}lp{0.80\textwidth}@{}}
\toprule
\textbf{Term} & \textbf{Definition} \\
\midrule
ARR & ACL Rolling Review submission cycle. \\
AR & Autoregressive (transformer decoder). \\
BPE & Byte-Pair Encoding (subword tokenization). \\
CER & Character Error Rate (ASR transcription error). \\
CI & 95\% Confidence Interval. \\
ECAPA-TDNN & Emphasized Channel Attention, Propagation and Aggregation TDNN. \\
EOS & End-of-sequence token. \\
FT & Fine-tuned (e.g., YourTTS (FT), fine-tuned on our corpus). \\
GT & Ground truth (natural human recording). \\
API & Application programming interface (commercial TTS services). \\
HF & Hugging Face. \\
HiFi-GAN & High-Fidelity GAN vocoder. \\
ITU-T P.808 & ITU recommendation for subjective speech-quality testing in remote conditions. \\
JSD & Jensen--Shannon divergence. \\
\textit{juktakkhor} & Bangla conjunct consonant cluster. \\
LLM & Large Language Model. \\
LoRA & Low-Rank Adaptation \cite{hu2022lora}. \\
LUFS & Loudness Units relative to Full Scale. \\
MOS & Mean Opinion Score. \\
NISQA & Non-Intrusive Speech Quality Assessment \cite{mittag2021nisqa}. \\
PII & Personally Identifiable Information. \\
RTF & Real-Time Factor (synth time / audio duration). \\
S1 / S2 / S3 & Three stages of the fine-tuning recipe. \\
SECS & Speaker Encoder Cosine Similarity. \\
SMOS & Speaker MOS (similarity-rated). \\
SNR & Signal-to-Noise Ratio. \\
T3 & Chatterbox's text-to-speech-token transformer (520M). \\
S3Gen & Chatterbox's speech-token-to-waveform vocoder. \\
TTR & Type-Token Ratio. \\
TTS & Text-to-Speech. \\
\bn{।} (\textit{dari}) & Bengali full-stop punctuation (U+0964). \\
VAD & Voice Activity Detection. \\
$\mathcal{L}_{\text{text}}$ & Next-token cross-entropy loss on text tokens (Eq.~\ref{eq:loss}). \\
$\mathcal{L}_{\text{speech}}^{\text{masked}}$ & Prompt-masked cross-entropy loss on speech tokens (Eq.~\ref{eq:loss_speech}). \\
$s_t$ & The $t$-th speech token (Eq.~\ref{eq:loss_speech}). \\
$p_\theta$ & Probability the T3 model (parameters $\theta$) assigns to the next token. \\
$T$ & Speech-token sequence length (Eq.~\ref{eq:loss_speech}). \\
$P_k$ & Unit histogram of the selected scripts at tier $k$ (Eq.~\ref{eq:jsd}). \\
$T_k$ & Target histogram for tier $k$: $T_k=\lambda P^{\text{nat}}_k+(1-\lambda)U_k$ (Eq.~\ref{eq:jsd}; \S\ref{sec:dataset}). \\
$\lambda$ & Target-interpolation weight between natural and uniform parts ($\lambda{=}0.7$). \\
$P^{\text{nat}}_k$, $U_k$ & Natural unit frequencies of the candidate pool; uniform distribution over the tier-$k$ inventory. \\
$w_k$ & Weight of tier $k$ in the tiered JSD objective (Eq.~\ref{eq:jsd}). \\
$L_{\text{prompt}}$ & Prompt-mask speech-token positions ($\approx 75$ at 3\,s). \\
$\mathbf{e}_{\text{spk}}$ & 256-d Voice-Encoder speaker embedding. \\
$\mathbf{c}_{\text{text}}$ & Normalized text-token context. \\
$|V|, |V_s|$ & Text vocab ($|V|{=}4{,}240$); speech vocab ($|V_s|{=}8{,}194$). \\
\bottomrule
\end{tabular}
\caption{Abbreviations, symbols, and Bangla terms used in the paper.}
\label{tab:abbrev}
\end{table*}


\begin{thebibliography}{41}
\providecommand{\natexlab}[1]{#1}

\bibitem[{Ahmad et~al.(2021)Ahmad, Selim, Iqbal, and Rahman}]{sust_tts_corpus}
Arif Ahmad, Md.~Reza Selim, Md.~Zafar Iqbal, and M.~Shahidur Rahman. 2021.
\newblock \href {https://doi.org/10.1250/ast.42.326} {{SUST TTS} corpus: A
  phonetically-balanced corpus for {Bangla} text-to-speech synthesis}.
\newblock \emph{Acoustical Science and Technology}, 42(6):326--332.
\newblock \textasciitilde30 hours, 17{,}357 utterances, single professional
  speaker, studio-quality; covers 1{,}208 Bangla phonetic units.

\bibitem[{Ansary(2024)}]{bnunicodenormalizer}
Nazmuddoha Ansary. 2024.
\newblock {bnUnicodeNormalizer}: {Bangla} {Unicode} normalization.
\newblock \url{https://github.com/mnansary/bnUnicodeNormalizer}.
\newblock Python library for normalizing {Bangla} {Unicode} text.

\bibitem[{Basher et~al.(2025)Basher, Kowsher, Islam, Nandi, Prottasha, Menon,
  Muntasir, Chowdhury, Alam, Yousefi, and Garibay}]{bntts2025}
Mohammad Jahid~Ibna Basher, Md~Kowsher, Md~Saiful Islam, Rabindra~Nath Nandi,
  Nusrat~Jahan Prottasha, Mehadi~Hasan Menon, Tareq~Al Muntasir, Shammur~Absar
  Chowdhury, Firoj Alam, Niloofar Yousefi, and Ozlem Garibay. 2025.
\newblock \href {https://doi.org/10.18653/v1/2025.findings-naacl.279} {{BnTTS}:
  Few-shot speaker adaptation in low-resource setting}.
\newblock In \emph{Findings of the Association for Computational Linguistics:
  {NAACL} 2025}, pages 4971--4983, Albuquerque, New Mexico. Association for
  Computational Linguistics.
\newblock Preprint: arXiv:2502.05729. Fine-tunes XTTS on \textasciitilde3.85k
  hours of Bangla speech; model/weights not publicly released.

\bibitem[{Casanova et~al.(2024)Casanova, Davis, G{\"o}lge, G{\"o}knar, Gulea,
  Hart, Aljafari, Meyer, Morais, Olayemi, and Weber}]{casanova2024xtts}
Edresson Casanova, Kelly Davis, Eren G{\"o}lge, G{\"o}rkem G{\"o}knar, Iulian
  Gulea, Logan Hart, Aya Aljafari, Joshua Meyer, Reuben Morais, Samuel Olayemi,
  and Julian Weber. 2024.
\newblock \href {https://doi.org/10.21437/Interspeech.2024-2016} {{XTTS}: a
  massively multilingual zero-shot text-to-speech model}.
\newblock In \emph{Proceedings of Interspeech}, pages 4978--4982.

\bibitem[{Casanova et~al.(2022)Casanova, Weber, Shulby, Junior, G{\"o}lge, and
  Ponti}]{casanova2022yourtts}
Edresson Casanova, Julian Weber, Christopher~D.\ Shulby, Arnaldo~Candido
  Junior, Eren G{\"o}lge, and Moacir~A.\ Ponti. 2022.
\newblock {YourTTS}: Towards zero-shot multi-speaker {TTS} and zero-shot voice
  conversion for everyone.
\newblock In \emph{Proceedings of the 39th International Conference on Machine
  Learning ({ICML})}, volume 162, pages 2709--2720.

\bibitem[{Chaudhury et~al.(2017)Chaudhury, Matin, Hossain, Uzzaman, and
  Masum}]{news_corpus_bangla}
Md. Tasnim~Haider Chaudhury, Abdul Matin, M.~S. Hossain, Asie Uzzaman, and Md.
  Masum. 2017.
\newblock Annotated {Bangla} news corpus and lexicon development with {POS}
  tagging and stemming.
\newblock \emph{Global Journal of Researches in Engineering}, 17(J1):5--12.
\newblock 74{,}698 word tokens from Prothom-Alo, Daily Janakantha, Daily
  Kalerkantho, and Amardesh; 13{,}550 distinct word forms annotated with POS,
  stemming, morphemes.

\bibitem[{Chowdhury and Adhikary(2022)}]{bnnumerizer}
Mahir~Labib Chowdhury and Aniruddha Adhikary. 2022.
\newblock bnnumerizer: {Bangla} numeral-to-word converter.
\newblock \url{https://github.com/banglakit/number-to-bengali-word}.
\newblock Python package for converting {Bengali} numerals to words.

\bibitem[{Dash et~al.(2009)Dash, Chowdhury, and Sarkar}]{collab_lexicon_bangla}
Niladri~Sekhar Dash, Payel~Dutta Chowdhury, and Abhishek Sarkar. 2009.
\newblock Naturalization of {English} words in modern {Bengali}: A corpus-based
  empirical study.
\newblock \emph{Language Forum}, 35(2):127--142.
\newblock Corpus-based analysis of English borrowing and naturalization of
  technical and scientific vocabulary in modern Bangla.

\bibitem[{Desplanques et~al.(2020)Desplanques, Thienpondt, and
  Demuynck}]{desplanques2020ecapa}
Brecht Desplanques, Jenthe Thienpondt, and Kris Demuynck. 2020.
\newblock \href {https://doi.org/10.21437/Interspeech.2020-2650} {{ECAPA-TDNN}:
  Emphasized channel attention, propagation and aggregation in {TDNN} based
  speaker verification}.
\newblock In \emph{Proceedings of Interspeech}, pages 3830--3834.

\bibitem[{Eberhard et~al.(2024)Eberhard, Simons, and Fennig}]{ethnologue2024}
David~M. Eberhard, Gary~F. Simons, and Charles~D. Fennig, editors. 2024.
\newblock \emph{Ethnologue: Languages of the World}, 27th edition.
\newblock SIL International, Dallas, Texas.
\newblock \url{https://www.ethnologue.com}.

\bibitem[{Ghosh et~al.(2024)Ghosh, Patra, and Dey}]{ghosh2024textnorm}
Krishnendu Ghosh, Munmun Patra, and Eshita Dey. 2024.
\newblock \href {https://doi.org/10.1007/978-3-031-78495-8\_27} {Enhancing
  {Bengali} text-to-speech synthesis through transformer-driven text
  normalization}.
\newblock In \emph{Proceedings of the International Conference on Pattern
  Recognition ({ICPR})}, pages 429--444. Springer.

\bibitem[{Gutkin et~al.(2016)Gutkin, Ha, Jansche, Pipatsrisawat, and
  Sproat}]{gutkin2016}
Alexander Gutkin, Linne Ha, Martin Jansche, Knot Pipatsrisawat, and Richard
  Sproat. 2016.
\newblock {TTS} for low resource languages: {A} {Bangla} synthesizer.
\newblock In \emph{Proceedings of the Tenth International Conference on
  Language Resources and Evaluation ({LREC})}, pages 2005--2010, Portoro\v{z},
  Slovenia.

\bibitem[{Hu et~al.(2022)Hu, Shen, Wallis, Allen-Zhu, Li, Wang, Wang, and
  Chen}]{hu2022lora}
Edward~J.\ Hu, Yelong Shen, Phillip Wallis, Zeyuan Allen-Zhu, Yuanzhi Li, Shean
  Wang, Lu~Wang, and Weizhu Chen. 2022.
\newblock {LoRA}: Low-rank adaptation of large language models.
\newblock In \emph{Proceedings of the International Conference on Learning
  Representations ({ICLR})}.

\bibitem[{{International Telecommunication Union}(2021)}]{itu_p808}
{International Telecommunication Union}. 2021.
\newblock \href {https://www.itu.int/rec/T-REC-P.808} {{ITU-T} recommendation
  {P.808}: Subjective evaluation of speech quality with a crowdsourcing
  approach}.
\newblock Technical report, ITU-T.

\bibitem[{Jahan et~al.(2015)Jahan, Kulsum, and Naser}]{sust_duration}
Labiba Jahan, Umme Kulsum, and Abu Naser. 2015.
\newblock {Bengali} diphone duration modeling for {Bengali} text-to-speech
  synthesis system.
\newblock \emph{American Academic \& Scholarly Research Journal}, 7(3).
\newblock SUBACHAN synthesizer; 527 diphones computed from 6 vowels and 32
  consonants.

\bibitem[{Kharitonov et~al.(2023)Kharitonov, Vincent, Borsos, Marinier, Girgin,
  Pietquin, Sharifi, Tagliasacchi, and Zeghidour}]{kharitonov2023speak}
Eugene Kharitonov, Damien Vincent, Zal{\'a}n Borsos, Rapha{\"e}l Marinier,
  Sertan Girgin, Olivier Pietquin, Matt Sharifi, Marco Tagliasacchi, and Neil
  Zeghidour. 2023.
\newblock Speak, read and prompt: High-fidelity text-to-speech with minimal
  supervision.
\newblock \emph{Transactions of the Association for Computational Linguistics},
  11:1703--1718.

\bibitem[{Kirkland et~al.(2023)Kirkland, Mehta, Lameris, Henter, Szekely, and
  Gustafson}]{kirkland2023mos}
Ambika Kirkland, Shivam Mehta, Harm Lameris, Gustav~Eje Henter, Eva Szekely,
  and Joakim Gustafson. 2023.
\newblock \href {https://doi.org/10.21437/SSW.2023-7} {Stuck in the {MOS} pit:
  A critical analysis of {MOS} test methodology in {TTS} evaluation}.
\newblock In \emph{12th ISCA Speech Synthesis Workshop ({SSW} 2023)}, pages
  41--47.

\bibitem[{Kong et~al.(2020)Kong, Kim, and Bae}]{kong2020hifigan}
Jungil Kong, Jaehyeon Kim, and Jaekyoung Bae. 2020.
\newblock {HiFi-GAN}: Generative adversarial networks for efficient and high
  fidelity speech synthesis.
\newblock In \emph{Proceedings of Advances in Neural Information Processing
  Systems ({NeurIPS})}.

\bibitem[{Kudo and Richardson(2018)}]{kudo2018sentencepiece}
Taku Kudo and John Richardson. 2018.
\newblock {SentencePiece}: A simple and language independent subword tokenizer
  and detokenizer for neural text processing.
\newblock In \emph{Proceedings of the 2018 Conference on Empirical Methods in
  Natural Language Processing: System Demonstrations}, pages 66--71, Brussels,
  Belgium. Association for Computational Linguistics.

\bibitem[{Kumar et~al.(2025)Kumar, Kumar, Sathe, and Pati}]{vits_bangla2025}
Sujeet Kumar, Siddharth Kumar, Kushal Sathe, and Jayadeep Pati. 2025.
\newblock \href {https://doi.org/10.1007/s10791-025-09701-3} {Advancing
  {Bangla} text-to-speech synthesis using a {VITS}-based model with a custom
  dataset and comprehensive evaluation}.
\newblock \emph{Discover Computing}, 28(1).
\newblock Article 183.

\bibitem[{Le et~al.(2023)Le, Vyas, Shi, Karrer, Sari, Moritz, Williamson,
  Manohar, Adi, Mahadeokar, and Hsu}]{le2023voicebox}
Matthew Le, Apoorv Vyas, Bowen Shi, Brian Karrer, Leda Sari, Rashel Moritz,
  Mary Williamson, Vimal Manohar, Yossi Adi, Jay Mahadeokar, and Wei-Ning Hsu.
  2023.
\newblock {Voicebox}: Text-guided multilingual universal speech generation at
  scale.
\newblock In \emph{Proceedings of Advances in Neural Information Processing
  Systems ({NeurIPS})}.

\bibitem[{Li et~al.(2023)Li, Han, Raghavan, Mischler, and
  Mesgarani}]{li2023styletts2}
Yinghao~Aaron Li, Cong Han, Vinay Raghavan, Gavin Mischler, and Nima Mesgarani.
  2023.
\newblock {StyleTTS} 2: Towards human-level text-to-speech through style
  diffusion and adversarial training with large speech language models.
\newblock In \emph{Proceedings of Advances in Neural Information Processing
  Systems ({NeurIPS})}.

\bibitem[{Loshchilov and Hutter(2019)}]{loshchilov2019adamw}
Ilya Loshchilov and Frank Hutter. 2019.
\newblock Decoupled weight decay regularization.
\newblock In \emph{Proceedings of the International Conference on Learning
  Representations ({ICLR})}.

\bibitem[{Mittag et~al.(2021)Mittag, Naderi, Chehadi, and
  M{\"o}ller}]{mittag2021nisqa}
Gabriel Mittag, Babak Naderi, Assmaa Chehadi, and Sebastian M{\"o}ller. 2021.
\newblock {NISQA}: {A} deep {CNN-Self-Attention} model for multidimensional
  speech quality prediction with crowdsourced datasets.
\newblock In \emph{Proc.\ Interspeech}, pages 2127--2131.

\bibitem[{Mukherjee and Mandal(2012)}]{mukherjee2012hmm}
Sankar Mukherjee and Shyamal Kumar~Das Mandal. 2012.
\newblock A {Bengali} {HMM} based speech synthesis system.
\newblock In \emph{Proceedings of the 2012 {Oriental} {COCOSDA} International
  Conference on Speech Database and Assessments}, pages 255--259, Macau, China.

\bibitem[{Nagarajan(2014)}]{loanwords_bangla}
Hemalatha Nagarajan. 2014.
\newblock Constraints through the ages: Loanwords in {Bangla}.
\newblock \emph{The {EFL} Journal}, 5(1):41--63.
\newblock Optimality-theoretic analysis of Sanskrit, Persian, Arabic,
  Portuguese, and English loanwords in Bangla; constraint re-ranking from
  Faithfulness to Markedness over time.

\bibitem[{{Narakeet}(2024)}]{narakeet}
{Narakeet}. 2024.
\newblock Narakeet: Text to speech online.
\newblock \url{https://www.narakeet.com}.
\newblock Accessed 2025.

\bibitem[{Pavel et~al.(2006)Pavel, Sarkar, Shah, and Khan}]{lexicon_bangla}
Dewan Shahriar~Hossain Pavel, Asif~Iqbal Sarkar, Faisal~Muhammad Shah, and
  Mumit Khan. 2006.
\newblock Collaborative lexicon development for {Bangla}.
\newblock In \emph{Proceedings of the Conference on Language and Technology},
  Dhaka, Bangladesh. Center for Research on Bangla Language Processing, {BRAC}
  University.
\newblock Annotated Bangla lexicon with POS, syntax, semantic and morphological
  tags.

\bibitem[{Praveen et~al.(2025)Praveen, Anand, Siddhartha, and Khapra}]{indicf5}
S~V Praveen, Srija Anand, Soma Siddhartha, and Mitesh~M. Khapra. 2025.
\newblock {IndicF5}: High-quality text-to-speech for {Indian} languages.
\newblock \url{https://github.com/AI4Bharat/IndicF5}.
\newblock AI4Bharat. Builds on F5-TTS.

\bibitem[{Radford et~al.(2023)Radford, Kim, Xu, Brockman, McLeavey, and
  Sutskever}]{whisper2023}
Alec Radford, Jong~Wook Kim, Tao Xu, Greg Brockman, Christine McLeavey, and
  Ilya Sutskever. 2023.
\newblock Robust speech recognition via large-scale weak supervision.
\newblock In \emph{Proceedings of the 40th International Conference on Machine
  Learning ({ICML})}, volume 202, pages 28492--28518. PMLR.

\bibitem[{Raju et~al.(2019)Raju, Bhattacharjee, Ahmad, and
  Rahman}]{sust_dnn_tts}
Rajan~Saha Raju, Prithwiraj Bhattacharjee, Arif Ahmad, and Mohammad~Shahidur
  Rahman. 2019.
\newblock \href {https://doi.org/10.1109/ICBSLP47725.2019.202055} {A {Bangla}
  text-to-speech system using deep neural networks}.
\newblock In \emph{Proceedings of the International Conference on Bangla Speech
  and Language Processing ({ICBSLP})}, pages 1--5. IEEE.
\newblock DNN-based statistical parametric TTS; 40h single-speaker corpus,
  135k-word lexicon.

\bibitem[{Rashid et~al.(2009)Rashid, Hussain, and Rahman}]{sust_diphone}
Muhammad~Masud Rashid, Md.~Akter Hussain, and M.~Shahidur Rahman. 2009.
\newblock Diphone preparation for {Bangla} text-to-speech synthesis.
\newblock In \emph{Proceedings of the 12th International Conference on Computer
  and Information Technology ({ICCIT})}, pages 226--230, Dhaka, Bangladesh.
  IEEE.
\newblock Concatenation-based synthesis with NLP (text normalization) and DSP
  (G2P) modules.

\bibitem[{{Resemble AI}(2025)}]{chatterboxtts2025}
{Resemble AI}. 2025.
\newblock {Chatterbox TTS}.
\newblock \url{https://github.com/resemble-ai/chatterbox}.
\newblock MIT License. 500M-parameter Llama-style autoregressive TTS with
  built-in zero-shot voice cloning.

\bibitem[{Sennrich et~al.(2016)Sennrich, Haddow, and Birch}]{sennrich2016bpe}
Rico Sennrich, Barry Haddow, and Alexandra Birch. 2016.
\newblock Neural machine translation of rare words with subword units.
\newblock In \emph{Proceedings of the 54th Annual Meeting of the Association
  for Computational Linguistics (Volume 1: Long Papers)}, pages 1715--1725,
  Berlin, Germany. Association for Computational Linguistics.

\bibitem[{Shen et~al.(2024)Shen, Ju, Tan, Liu, Leng, He, Qin, Zhao, and
  Bian}]{shen2024naturalspeech2}
Kai Shen, Zeqian Ju, Xu~Tan, Eric Liu, Yichong Leng, Lei He, Tao Qin, Sheng
  Zhao, and Jiang Bian. 2024.
\newblock {NaturalSpeech} 2: Latent diffusion models are natural and zero-shot
  speech and singing synthesizers.
\newblock In \emph{Proceedings of the International Conference on Learning
  Representations ({ICLR})}.

\bibitem[{{Silero Team}(2024)}]{silero_vad}
{Silero Team}. 2024.
\newblock Silero {VAD}: Pretrained enterprise-grade voice activity detector
  ({VAD}), number detector and language classifier.
\newblock \url{https://github.com/snakers4/silero-vad}.
\newblock Used at inference time for per-chunk tail-silence trimming.

\bibitem[{Sinha et~al.(2012)Sinha, Jana, Dasgupta, and
  Basu}]{semantic_lexicon_bangla}
Manjira Sinha, Abhik Jana, Tirthankar Dasgupta, and Anupam Basu. 2012.
\newblock \href {https://aclanthology.org/W12-5114/} {A new semantic lexicon
  and similarity measure in {Bangla}}.
\newblock In \emph{Proceedings of the 3rd Workshop on Cognitive Aspects of the
  Lexicon (CogALex)}, pages 171--182, Mumbai, India. Association for
  Computational Linguistics.

\bibitem[{Thienpondt and Demuynck(2023)}]{thienpondt2023ecapa2}
Jenthe Thienpondt and Kris Demuynck. 2023.
\newblock {ECAPA2}: A hybrid neural network architecture and training strategy
  for robust speaker embeddings.
\newblock In \emph{Proceedings of the {IEEE} Automatic Speech Recognition and
  Understanding Workshop ({ASRU})}. IEEE.

\bibitem[{Uddin et~al.(2015)Uddin, Sakib, Rupu, Hossain, and
  Huda}]{sust_phoneme}
Mir~Ashraf Uddin, Nazmus Sakib, Esrat~Farjana Rupu, Md.~Afzal Hossain, and
  Md.~Nurul Huda. 2015.
\newblock Phoneme based {Bangla} text to speech conversion.
\newblock In \emph{Proceedings of the 18th International Conference on Computer
  and Information Technology ({ICCIT})}, pages 531--533. IEEE.
\newblock Phoneme-based synthesis; recording and separation of Bangla phonemes
  for improved speech quality.

\bibitem[{Wang et~al.(2025)Wang, Chen, Wu, Zhang, Zhou, Liu, Chen, Liu, Wang,
  Li, He, Zhao, and Wei}]{wang2023valle}
Chengyi Wang, Sanyuan Chen, Yu~Wu, Ziqiang Zhang, Long Zhou, Shujie Liu, Zhuo
  Chen, Yanqing Liu, Huaming Wang, Jinyu Li, Lei He, Sheng Zhao, and Furu Wei.
  2025.
\newblock \href {https://doi.org/10.1109/TASLPRO.2025.3530270} {Neural codec
  language models are zero-shot text to speech synthesizers}.
\newblock \emph{IEEE Transactions on Audio, Speech and Language Processing},
  33:705--718.

\bibitem[{Yang et~al.(2025)Yang, Wang, Han, Liu, Li, Qin, and
  Chen}]{yang2025responsibleeval}
Yifan Yang, Hui Wang, Bing Han, Shujie Liu, Jinyu Li, Yong Qin, and Xie Chen.
  2025.
\newblock \href {https://arxiv.org/abs/2510.06927} {Position: Towards
  responsible evaluation for text-to-speech}.
\newblock \emph{Preprint}, arXiv:2510.06927.

\end{thebibliography}
\end{document}